\makeatletter
\disable@package@load{program}{}
\makeatother

\documentclass[pdflatex,sn-mathphys-num]{sn-jnl}

\makeatletter
\let\orcidlogo\@undefined
\makeatother

\usepackage{orcidlink}

\usepackage{graphicx}
\usepackage{tabularx}
\usepackage{multirow}
\usepackage{amsmath,amssymb,amsfonts}
\usepackage{amsthm}%
\usepackage{mathrsfs}%
\usepackage[title]{appendix}%
\usepackage{textcomp}%
\usepackage{manyfoot}%
\usepackage{booktabs}%
\usepackage{algorithm}%
\usepackage{algorithmicx}%
\usepackage{algpseudocode}%
\usepackage{listings}%
\usepackage{mhchem}
\usepackage{braket}
\usepackage{soul}
\usepackage{bm}
\usepackage{subcaption}
\usepackage{multirow, multicol}
\usepackage{makecell}

\usepackage{booktabs}
\usepackage{caption}
\usepackage{siunitx}

\usepackage{xcolor}
\usepackage{listings}

\lstdefinelanguage{XYZ}{
  morecomment=[l]{\#},
  morestring=[b]",
}

\lstdefinestyle{xyzstyle}{
  basicstyle=\ttfamily\footnotesize,
  columns=fullflexible,
  frame=single,
  rulecolor=\color{black!25},
  backgroundcolor=\color{black!3},
  xleftmargin=1em,
  xrightmargin=1em,
  framexleftmargin=0.75em,
  framexrightmargin=0.75em,
  breaklines=false,
  numbers=none,
  showstringspaces=false,
  aboveskip=0.75em,
  belowskip=0.75em
}

\usepackage{xcolor}

\newcommand{\opcoeffpair}{``\texttt{operator coefficient}'' }

\theoremstyle{thmstyleone}%

\theoremstyle{thmstyletwo}%

\theoremstyle{thmstylethree}%

\begin{document}

\title[ai4chem]{Learning to Prepare Molecular Ground States with Transformer Models}

\author[1]{Alex Koziell-Pipe\,\orcidlink{0000-0003-4773-3589}}
\equalcont{These authors contributed equally to this work.}\email{Alex.Koziell-Pipe@quantinuum.com}
\author[1]{Jasmine Brewer\,\orcidlink{0000-0002-3084-0663}}
\equalcont{These authors contributed equally to this work.}
\author[1]{Jem Guhit\,\orcidlink{0000-0002-9802-0901}}
\equalcont{These authors contributed equally to this work.}
\author[3]{Marwa H. Farag}
\equalcont{These authors contributed equally to this work.}\email{mfarag@nvidia.com}
\author[2]{Kripa Panchagnula\,\orcidlink{0009-0004-3952-073X}}
\author[2]{Gabriel Laude\,\orcidlink{0000-0002-1373-9504}}
\author[1]{Fabian Finger\,\orcidlink{0000-0002-6996-5957}}
\author[2]{Carlo Gaggioli\,\orcidlink{0000-0001-9105-8731}}
\author[2]{Ludmila Szulakowska\,\orcidlink{0000-0002-0188-2207}}
\author[2]{Oliver J. Backhouse\,\orcidlink{0000-0002-3086-1704}}
\author[5]{Christos Papalitsas\,\orcidlink{0000-0003-0467-796X}}
\author[4]{Jason G. Mustakis}
\author[2]{Thomas Soini\,\orcidlink{0000-0003-3237-3509}}
\author[2]{David Mu\~{n}oz Ramo\,\orcidlink{0000-0002-7220-9466}}
\author[1]{Stephen Clark}
\author[3]{Elica Kyoseva\,\orcidlink{0000-0002-9154-0293}}
\author[1]{Enrico Rinaldi\,\orcidlink{0000-0003-4134-809X}}


\affil[1]{\orgname{Quantinuum}, \orgaddress{London, UK}}
\affil[2]{\orgname{Quantinuum}, \orgaddress{Cambridge, UK}}

\affil[3]{\orgname{NVIDIA Corporation}, \orgaddress{Santa Clara, CA, USA}}

\affil[4]{\orgdiv{Chemical R\&D}, \orgname{Pfizer Inc.}, \orgaddress{Groton, CT, USA}}

\affil[5]{\orgdiv{Center for Digital Innovation}, \orgname{Pfizer Inc.}, \orgaddress{Thessaloniki, Greece}}

\abstract{
Quantum state preparation is a key component of many quantum algorithms. Performing this step efficiently is essential for realizing practical quantum advantage in quantum chemistry applications. Iterative algorithms like ADAPT-VQE can produce shallow ground-state preparation circuits, but become computationally prohibitive for the larger molecules relevant to materials science and pharmaceutical development. Here, we introduce ADAPT-GQE, a generative AI framework that learns to synthesize ground-state preparation circuits for electronic structure calculations. 
We first use ADAPT-VQE to generate high-quality reference circuits, which are then used as targets for training models for circuit generation.
Once trained, the model can efficiently propose and score circuits, enabling reinforcement learning (RL) to drive circuit generation accuracy beyond the accuracy of the ADAPT-VQE training data. This pipeline achieves order-of-magnitude reductions in circuit generation time relative to ADAPT-VQE while maintaining comparable or improved state-preparation accuracy. We demonstrate ADAPT-GQE on imipramine, a well-established tricyclic antidepressant that serves as a representative, challenging target for computational modelling in drug stability protocols. We execute generated circuits on Quantinuum Helios-1, representing a milestone for AI-generated quantum chemistry circuits on state-of-the-art quantum hardware. These results establish a pathway toward automated quantum circuit synthesis for utility-scale quantum computational chemistry.
}




\maketitle

\section{Introduction}\label{introduction}

Quantum computers hold the promise of solving computational problems that are fundamentally intractable for classical hardware, with potential applications spanning combinatorial optimization, cryptography, materials discovery, and quantum chemistry~\cite{Farhi2014QAOA,Shor1997Algorithms,Cao2019QuantumChemistry}. A central primitive underlying many of these applications is quantum state preparation: the task of constructing a quantum circuit that drives a system into a target state, such as the ground state of a problem Hamiltonian. The quality and efficiency of state preparation directly determines the accuracy and resource cost of downstream quantum computations. In the domain of quantum chemistry — one of the compelling near-term applications — accurately computing molecular ground-state energies requires simulating quantum mechanical interactions that scale exponentially with system size on classical hardware, rendering exact methods such as Full Configuration Interaction (FCI) infeasible beyond small molecules~\cite{olsen_bible}. Quantum devices, by contrast, encode fermionic many-body states in qubits, and quantum algorithms such as Quantum Phase Estimation (QPE) offer polynomial-scaling solutions in principle~\cite{AspuruGuzik2005, McArdle:2018tza}. Yet realizing this advantage in practice remains a formidable challenge: present-day noisy intermediate-scale quantum (NISQ) devices are constrained by limited qubit counts, short coherence times, and high gate error rates~\cite{Preskillnisq}. Circuits must be kept shallow, optimization must tolerate noise, and algorithms must work within tight resource budgets — requirements that existing state preparation methods satisfy only partially, and which motivate the development of more efficient, automated approaches.

The Variational Quantum Eigensolver (VQE)~\cite{Peruzzo2014VQE,McClean2016Theory} has emerged as the leading paradigm for near-term quantum chemistry, offering a hybrid quantum-classical approach in which a parameterized quantum circuit (the ansatz) prepares a trial wavefunction and a classical optimizer minimizes the expected energy. 
VQE is appealing because, in principle, it can be implemented using relatively shallow parametrized circuits and exhibits some robustness to hardware noise. However, VQE is subject to several well-documented and severe limitations. First, in practice its performance depends critically on the choice of ansatz. Widely used chemistry-inspired ansätze, such as Unitary Coupled Cluster Singles and Doubles (UCCSD), often require circuit depths that grow rapidly with molecular size and can exceed the coherence limits of current quantum hardware~\cite{Cao2019QuantumChemistry}.
Second, the classical optimization of VQE is plagued by barren plateaus — regions of the parameter landscape where gradients vanish exponentially with the number of qubits — making convergence unreliable at scale~\cite{Cerezo2021VQA,McClean2018BarrenPlateaus}. Third, the number of variational parameters grows rapidly with system size, compounding shot-noise requirements and rendering the classical optimization increasingly expensive~\cite{McClean2016Theory,Grimsley2019ADAPTVQE}.

The Adaptive Derivative-Assembled Problem-Tailored Variational Quantum Eigensolver, ADAPT-VQE~\cite{Grimsley2019ADAPTVQE}, was introduced to address the circuit depth problem through a fundamentally different construction strategy. Rather than committing to a fixed ansatz \textit{a priori}, ADAPT-VQE iteratively grows the circuit by selecting, at each step, the operator from a predefined pool that results in the greatest energy decrease — as measured by the gradient. This construction enables compact, problem-specific ansätze that are tailored to the electronic structure of the target molecule, and typically yields significantly shallower circuits than fixed-structure approaches while maintaining high accuracy. Despite these advantages, ADAPT-VQE does not escape the fundamental scaling bottleneck of variational methods~\cite{Grimsley2019ADAPTVQE}. At every iteration, it must evaluate gradients for all operators in the pool, requiring a number of quantum circuit executions that grows as $O (\mathrm{pool\,size} \times d)$ where $d$ is the circuit depth — a cost that becomes prohibitive beyond systems of roughly 15 qubits. More critically, the full ADAPT-VQE procedure also requires global re-optimisation of all accumulated variational parameters at each step, adding an optimiser-dependent cost that increases with the number of selected operators. This cost is compounded by the fact that the procedure must be repeated independently for every new molecular geometry, without a natural way of reusing information across related molecular configurations.

These limitations suggest that ADAPT-VQE, while powerful, is unlikely to serve as a scalable stand-alone solution for quantum state preparation in chemically and industrially relevant regimes. This work addresses that bottleneck by using generative AI to learn reusable structure from previously constructed ground-state circuits and transfer that knowledge to new molecular configuration instances. Across a rapidly expanding range of scientific and technical domains, generative AI has shown an unprecedented ability to identify structure in highly complex search spaces and generate high-quality solutions beyond the reach of traditional hand-designed methods. We build on this capability by integrating generative models with adaptive variational algorithms, with the goal of accelerating the construction of compact, high-accuracy quantum circuits for molecular electronic-structure problems.

As an illustrative example, we consider the problem of preparing ground states of different physical configurations of imipramine, a tricyclic antidepressant of interest for pharmaceutical development. Accurate molecular modeling is essential in early-stage pharmaceutical discovery and preclinical development, but the electronic-structure calculations underlying these predictions become rapidly intractable on classical hardware for large, flexible, and chemically complex molecules. Imipramine has multiple chemically distinct reactive sites and significant conformational flexibility, making it both a challenging target for electronic-structure modeling and a representative test case for chemically complex molecular systems. Its rich conformational landscape provides a concrete setting in which to evaluate the ability of generative models to learn transferable structure across related geometries and use that information to guide quantum state preparation.

In this work, we introduce a generative AI framework that produces complete quantum circuits for preparing the ground state of imipramine in a single forward pass, without iterative optimization at inference time. We consider two model classes: a very small transformer trained from scratch on problem-specific data, and a larger language model pretrained on general-purpose data and fine-tuned for quantum circuit generation. Both models are trained for next-token prediction using high-quality ground-state preparation circuits generated by ADAPT-VQE. For the smaller transformer, we further demonstrate that RL can systematically improve circuit accuracy beyond that of the ADAPT-VQE training dataset, rather than merely imitating it. Across 12--16 qubit active spaces of imipramine, the generated circuits approach or exceed the accuracy of the ADAPT-VQE training data while reducing circuit generation time by $3$--$4$ orders of magnitude relative to ADAPT-VQE for the smaller model and exhibiting substantially more favorable scaling with system size. As a proof of concept, we execute generated circuits on state-of-the-art Quantinuum quantum hardware. Together, these results demonstrate a scalable framework for automated quantum circuit synthesis for a pharmaceutically relevant molecular system, producing circuits that meet or exceed state-of-the-art methods in a fraction of the time and are executable on existing quantum hardware.

The remainder of this article is structured as follows. In Section~\ref{sec:state-of-the-art} we briefly review the state-of-the-art and previous work. In Section~\ref{sec:methods} we describe the detailed architectures and training pipelines of the models used in this work. In Section~\ref{sec:results} we demonstrate the ability of the models to produce circuits preparing the ground state in terms of their energy accuracy, illustrate their dramatic computational advantages with respect to ADAPT-VQE, and execute representative generated circuits on Quantinuum Helios. We conclude with Discussion and Conclusions in Sections~\ref{sec:discussion} and \ref{sec:conclusion}.

\section{State-of-the-Art and Previous Work}
\label{sec:state-of-the-art}

\subsection{AI for quantum state preparation}

The limitations of VQE and ADAPT-VQE have spurred growing interest in machine learning and artificial intelligence as a means of mitigating the computational bottlenecks of variational quantum algorithms and improving their scalability to larger and more complex systems.
Several distinct paradigms have been explored in the literature. One prominent line of work applies RL to quantum circuit design: an agent learns to construct or optimize gate sequences guided by an energy or fidelity reward signal \cite{Krumtunger2025RLQuantumCircuitArchitectures,Rieckmann2025GateSequenceRL}. Krumt\"unger et al.\ demonstrate an RL framework that learns problem-dependent quantum circuit mappings for ground-state preparation of molecular Hamiltonians across varying molecular geometries, while Rieckmann et al. use RL to optimize entangling gate sequences in quantum state preparation, achieving higher fidelity at reduced CNOT count compared to hardware-efficient ans\"atze. 

Rather than iteratively constructing the ansatz as in ADAPT-VQE, a second paradigm uses generative models to learn and generate problem-specific circuits. 
The Generative Quantum Eigensolver (GQE) trains a generative pre-trained transformer (GPT) to produce operator sequences conditioned on a molecular Hamiltonian. In contrast to fixed-ansatz VQE approaches, where optimization is performed over variational parameters within a predefined quantum circuit, GQE places the trainable parameters in a classical generative model and updates them online using energy feedback from the quantum processor.
Conditional-GQE \cite{Minami2025ConditionalGQE} extends this framework with an encoder-decoder architecture that enables generalization to new Hamiltonian instances without retraining. Generative Flow Networks have also been applied to circuit synthesis \cite{Dai2025FlowQNet}, achieving 10–30× more compact circuits than standard unitary baselines. 

A third direction applies large language models to quantum algorithm generation: QAOA-GPT~\cite{Tyagin2025QAOAGPT}  introduces a GPT-based framework that synthesizes adaptive QAOA circuits for combinatorial optimization problems from synthetic training data generated with adaptive QAOA in the CUDA-Q platform~\cite{cudaq, cudaq_github}. The model learns to generate high-quality, problem-specific circuits for unseen instances without iterative gradient evaluation, demonstrating the potential of generative transformers for automated quantum circuit synthesis. Finally, parameter transfer and warm-starting strategies exploit the smoothness of quantum energy landscapes across related problem instances. Xu et al.~\cite{Xu2025QAOAParameterTransfer} use Graph Attention Networks to transfer optimized QAOA parameters across graph instances; Bincoletto et al.~\cite{Bincoletto2025TransferableVQE} develop transferable machine learning models that predict VQE circuit parameters across different molecules, demonstrating systematic transferability to systems larger than those seen in training; and Qracle~\cite{Zhang2025Qracle} uses a graph neural network to encode both the Hamiltonian and ansatz circuit structure, learning a joint mapping that accelerates VQE convergence and improves final energy accuracy. Despite this broad progress, existing approaches share critical limitations: most are evaluated on small and structurally-homogeneous problem instances without addressing the challenge of conformer-level generalization in realistic molecular systems, and none has combined GPU-accelerated offline data generation with inference-time circuit synthesis and validated the resulting circuits on real quantum hardware.

\subsection{Pharmaceutical use-case}

The chemical degradation of active pharmaceutical ingredients (APIs) is a critical consideration throughout pharmaceutical development, as it directly impacts drug safety and efficacy.
Predicting the shelf-life of formulations under real storage conditions requires understanding the stability in solid state and considering API excipient interactions.
As experimental studies of the chemical kinetics of organic molecules in solid state are difficult, series of forced degradation studies are typically used as surrogates.
During this forced degradation, harsh chemical environments are utilized to push the drug to extreme limits.
The most common techniques include acid/base hydrolysis, oxidation and photolysis.
In order to understand the possible oxidative product and pathways free radical initiators and transition metal ions are used.
Understanding the rate of formation under these conditions then allows the extrapolation of possible degradation products under the more mild storage conditions.

While experimental stress testing remains the regulatory standard, it is often time-consuming, resource-intensive, and suffers from limited API availability during early development stages~\cite{Wu2022MolPharm}.
For this reason, first-principles computational methods have been developed to complement experimental work and predict API degradation pathways and kinetics directly from molecular structure.
Combined with automated reaction mechanism exploration such computational methods have been shown to enable predictive kinetic models for free-radical oxidative degradation in the presence of a widely-used azobisisobutyronitrile (AIBN) initiator~\cite{GrinbergDana2021MolPharm,AIChEImipramine2021}. 
While classical ab initio electronic structure calculations can successfully complement traditional experimental workflows, the complexity of the methodology and the steep computational cost reduces its practicality.
This is particularly the case for the description of static correlation effects expected to play a significant role in complex radical systems.

\section{Methods}\label{methods}
\label{sec:methods}

This section describes the key components of developing language models for ground state circuit generation. Section~\ref{subsec:DataGen} describes training dataset construction, including generating diverse conformational structures of imipramine and calculating high-quality energy-minimizing circuits for each conformer using ADAPT-VQE. Section~\ref{subsec:ml-methods} discusses training of our models, including how quantum circuits are represented as inputs and outputs for our transformer architectures (\ref{subsubsec:tokenization}),  a strategy for encoding molecular information as multimodal input (\ref{subsubsec:multimodality}), and architecture and training specifics for the two distinct models we consider (\ref{subsubsec:gemma3-training} and \ref{subsubsec:nemotron-training}). In particular, the architecture and training sections detail two complementary transformer architectures used in this work: a compact Gemma~3-based model~\cite{gemmateam2025gemma3technicalreport} with no prior pretraining, and a larger Nemotron~2~Nano~\cite{nvidia2025nvidianemotronnano2} model pretrained on a general-purpose dataset, together with the training pipelines used to adapt them for circuit generation.

\subsection{Data Generation}\label{subsec:DataGen}

We generate data according to two requirements aligned with real-world application. First, the dataset must broadly cover the accessible conformational space of imipramine to ensure that training and evaluation occur across a set of conformers representing the diversity of pharmaceutical use cases. Second, the dataset must contain a structurally well-separated subset that can be held out for testing: this enables evaluation of model generalization to a regime where conformers are not close structural analogues of those seen during training. To meet these requirements, we generate a diverse ensemble of imipramine conformers using a combination of molecular dynamics simulations, nudged elastic band calculations, and targeted manual perturbations:
\begin{itemize}
\item \textbf{Molecular dynamics (MD) ---} The bulk of the dataset is comprised of conformer snapshots from MD simulations. They provide a natural way to sample thermally accessible conformations and generate diverse geometries while preserving realistic intramolecular correlations. Simulations are performed using the OpenMM MD engine~\cite{openmm} coupled with the MACE-OFF potential~\cite{mace-off}. Dynamics are propagated in the NVT (canonical) ensemble using a Langevin thermostat, with initial velocities sampled from a Maxwell–Boltzmann distribution at 360 K and a friction coefficient of 1 $\mathrm{ps}^{-1}$. We run five independent trajectories of 20 million steps each with a timestep of 0.25 fs, and subsample conformer geometries from these simulations. The long trajectories ensure sufficient decorrelation between sampled configurations and reduce the likelihood of selecting nearly identical conformers.
\item \textbf{Nudged elastic band (NEB) transition pathways ---} 
MD simulations can undersample rare transitions and fail to explore conformational regions that are kinetically inaccessible from the initial conditions on accessible simulation timescales.
To ensure diversity we additionally calculate reference conformers of imipramine using the ETKDG algorithm implemented in RDKit \cite{landrum2006rdkit}, which produces sets of chemically plausible geometries that approximate distinct regions of conformational space. For more and less aggressive parameter choices these give rise to sets of 15 and 122 reference conformers. We augment the dataset with configurations along the transition pathways connecting 15 reference conformers. NEB calculations sample transition pathways between each pair of reference conformers. NEB pathways are computed with the Atomic Simulation Environment (ASE) package \cite{ase-paper} and each pathway is discretised into 30 intermediate images.
\item \textbf{Out-of-distribution (OoD) conformer perturbations ---} MD trajectories can contain highly similar configurations, either because nearby snapshots are strongly correlated or because the simulation repeatedly visits the same regions of conformational space. To enable a more stringent generalization test in which structurally similar conformers are not shared between training and test sets, we additionally construct a set of configurations designed to be well separated from the MD and NEB-derived structures. Starting from the highest-energy conformer among 122 reference conformers, we generate new geometries by perturbing the nuclear positions using the rattle method with a standard deviation of 0.05 
\AA. This procedure yields a set of high-energy conformers that are energetically separated from those obtained through MD and transition-pathway sampling.
\end{itemize}
Further details on the dataset generation are described in Appendix~\ref{appendix:molecular_data_generation}.

\begin{figure}[htbp]
    \centering
    \includegraphics[width=\linewidth]{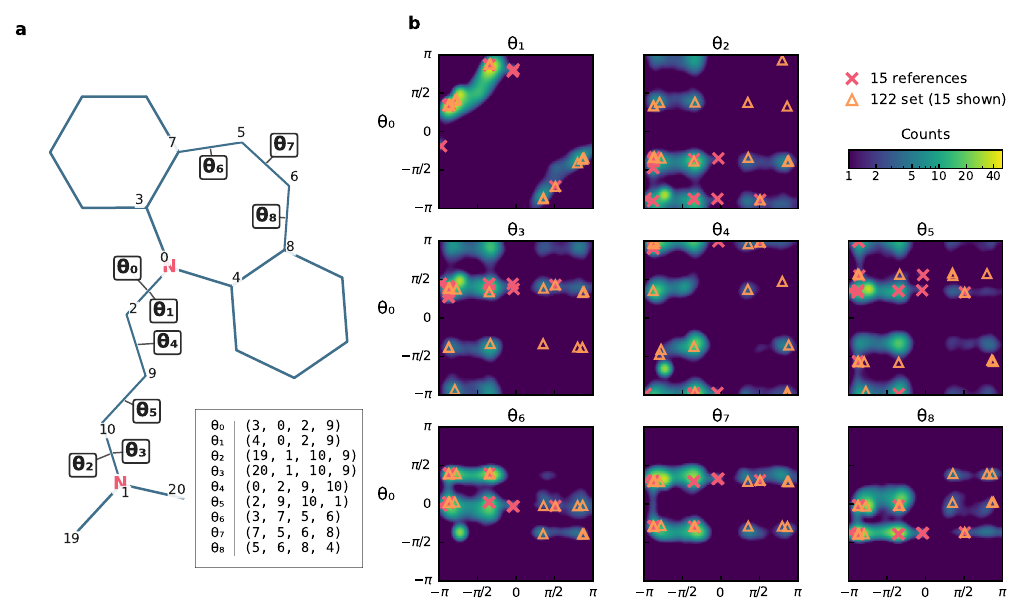}
    \caption{\textbf{Imipramine dihedral definitions and conformational density.}
    \textbf{(a)} Imipramine structure with the nine selected dihedral angles labelled. Atom numbers indicate the indices used to define each dihedral angle, and the inset table lists the corresponding four-atom tuples. \textbf{(b)} Dataset density in dihedral-angle space for $\theta_0$ against each remaining selected dihedral angle $\theta_1$--$\theta_8$. Color indicates smoothed dataset counts for $(\theta_0, \theta_i)$  values on a logarithmic scale. Pink crosses show the 15-reference conformer set; orange open triangles show the displayed subset from the 122-reference conformer set. The dataset forms well-separated clusters across the allowed dihedral-angle space, broadly consistent with the reference conformers.}
    \label{fig:data_diversity}
\end{figure}

We find that this procedure successfully generates a diverse set of conformational structures of imipramine. We characterize structures of imipramine through a set of 9 dihedral angles, including 6 non-ring dihedral angles and 3 angles characterizing the central ring of imipramine (see Figure~\ref{fig:data_diversity}a). As a representative example, we show in Figure~\ref{fig:data_diversity}b the marginal probability distribution over the dataset for the correlation of a typical dihedral angle $\theta_0$ with the other 8 dihedral angles. The dataset forms high-density clusters that span the full available regions of all dihedral angles and indicate good conformational diversity. These high-density regions also roughly correspond to the reference structures calculated with the ETKDG algorithm, which are overlaid as markers. The full marginal probability distributions for the remaining 7 dihedral angles are similar and are provided in Appendix~\ref{appendix_sub:data_diversity}.

For each conformational structure of imipramine obtained from the procedures described above, we apply ADAPT-VQE to generate compact quantum circuits that prepare approximations to the electronic ground state. ADAPT-VQE is a variational algorithm that iteratively constructs a system-specific ansatz by selecting operators from a predefined pool, and can achieve high accuracy with significantly fewer parameters than fixed-structure approaches. ADAPT-VQE is implemented using the CUDA-Q platform --- an open-source, qubit-agnostic platform for hybrid quantum--classical computing with GPU-accelerated simulation capabilities --- leveraging its GPU-accelerated state-vector simulator to efficiently execute large-scale quantum circuits. For a 12-qubit system, we observe a 234× speedup, reducing runtime to 43.4 seconds on a single NVIDIA H100 GPU, compared to over 10,000 seconds on a single CPU core (Intel Xeon Platinum 8480CL), demonstrating the substantial acceleration enabled by GPU-based simulation. For the 16-qubit case, distributing across eight H100 GPUs yields an additional 4.7× speedup relative to the single-GPU configuration (see Table~\ref{tab:ADAPT-VQE-scaling} in Appendix~\ref{appendix:compute} for detailed results).

To assess the capability and scalability of generative models for ground-state circuit synthesis across different system sizes, we generate datasets for each conformer with active spaces corresponding to 12, 14, and 16 qubits. The active-space electronic Hamiltonians are constructed using molecular integrals generated with PySCF and the 6-31G basis set ~\cite{Sun2018PySCF,Sun2020PySCF,Hehre1972GaussianBasis}. One- and two-electron integrals are computed from Hartree–Fock (HF) molecular orbitals for active-space configurations chosen to match the target qubit counts: a (6e,6o) active space for 12 qubits, (6e,7o) for 14 qubits, and (8e,8o) for 16 qubits. The resulting second-quantized fermionic Hamiltonians are mapped to qubit operators via the Jordan–Wigner transformation~\cite{Jordan1928Wigner,Tranter2018JordanWigner}.

The operator pool selected for ADAPT-VQE controls the trade-off between circuit expressivity, depth, and gradient evaluation cost. The UCCSD pool consists of singles and doubles fermionic excitation operators constrained by the occupied–virtual orbital structure of a single-determinant reference, while the UCCGSD pool generalizes these to all spin-orbital pairs without occupied–virtual restrictions~\cite{Lee2019GeneralizedUCC,Grimsley2019ADAPTVQE}, enabling more direct capture of multireference and strongly correlated electronic structure. We employ the UCCGSD pool for the 12- and 14-qubit cases, where the number of generalized excitations remains tractable and use the more compact UCCSD pool for the 16-qubit configuration to control computational cost while still generating chemically meaningful ansätze. We terminate ADAPT-VQE when the energy of the obtained circuit is within a fixed energy tolerance $\varepsilon$ of a reference energy. For 12- and 14-qubit systems we use the CASCI energy for reference; for the 16-qubit system we use the CCSD energy for consistency with the UCCSD ansatz used for this active space.
Thus each element of the dataset is a sequence of operators from the specified operator pool along with coefficients that, when applied to the Hartree-Fock state, would prepare the ground state of the specified configuration of imipramine to at least the accuracy specified by the tolerance.

We use five datasets in this work, corresponding to 12-, 14-, and 16-qubit active spaces. For the 12- and 14-qubit active spaces, we include ADAPT-VQE sequences generated with two energy-convergence thresholds: $5$ and $10$~mHa for the 12-qubit active space, $5$ and $16$~mHa for the 14-qubit active space, and $15$~mHa for the 16-qubit active space. The different thresholds allow us to distinguish molecular complexity from the circuit length required to reach a given accuracy.
The datasets have sizes ranging from approximately 13,000 to 15,700 circuits, and all datasets draw from all three conformer generation methods described above: MD structures, configurations from NEB transition pathways, and OoD perturbations. MD conformers constitute the majority in all cases (76--89\%), with NEB configurations accounting for 8--17\% and OoD perturbations for 4--7\%. Detailed characteristics of the five datasets are summarized in Table \ref{tab:datasets} of Appendix~\ref{appendix:molecular_data_generation}. 

\begin{figure}[t]
    \centering
    \includegraphics[width=0.95\textwidth]{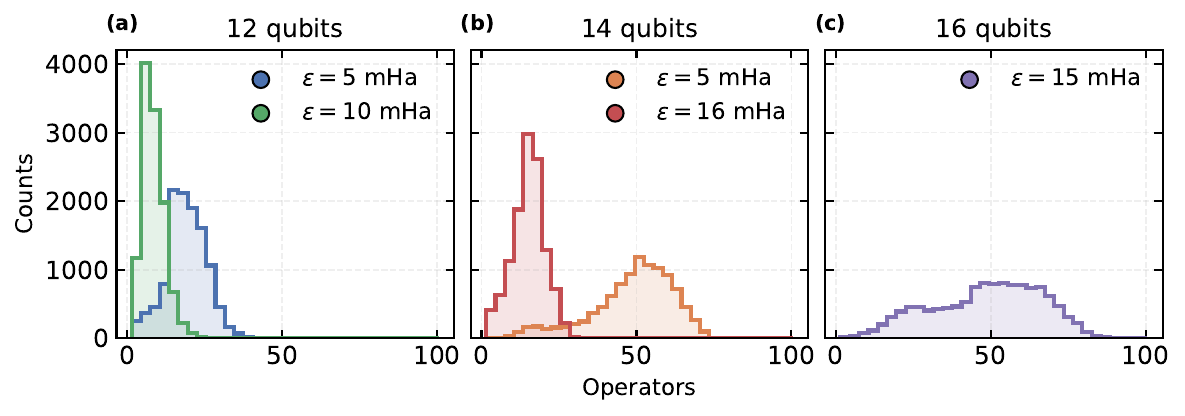}
\caption{\textbf{Operator counts for each dataset.} Number of operators from the operator pool for the 12- \textbf{(a)}, 14- \textbf{(b)}, and 16-qubit \textbf{(c)} systems in the ADAPT-VQE training data. The 12‑ and 14‑qubit cases use a UCCGSD pool with energy thresholds defined as the difference from CASCI, while the 16‑qubit case uses a UCCSD pool with error measured relative to CCSD.}
    \label{fig:circuit-depth}
\end{figure}

The number of operators selected by ADAPT-VQE during circuit generation, and the resulting circuit depth, vary substantially both for the chosen active space and the energy convergence threshold, see Figure~\ref{fig:circuit-depth}. Higher-accuracy circuits require more operators and deeper circuits: for the 12 qubit systems, tightening the threshold from $\varepsilon = 10$ mHa to $\varepsilon = 5$ mHa roughly doubles the median operator count (from 8 to 19 UCCGSD operators). For the 14 qubit system the stricter threshold ($\varepsilon = 5$ mHa) yields a median of 51 operators compared to 15 operators at the looser threshold ($\varepsilon = 16$ mHa). The 16 qubit system at a threshold of $\varepsilon = 15$ mHa has a median operator count of $50$, similar to the 14-qubit system with low tolerance. Increasing the active-space size leads to a broader spread in operator count, consistent with greater variability among molecular conformers in the larger active spaces.

\subsection{Circuit Generation with Language Models}
\label{subsec:ml-methods}


The molecular Hamiltonians and ADAPT-VQE circuits generated in Section~\ref{subsec:DataGen} form the basis of our training data. Conditioned on information about the molecular Hamiltonian, we train language models to generate circuits that minimize the corresponding state energy. The ADAPT-VQE circuits provide reference sequences for training, during which the models learn to reproduce the statistical structure of ADAPT-VQE-generated circuits. The same Hamiltonian information is used as conditioning input during RL, where the model is optimized directly as a policy for generating low-energy circuits.

We train two complementary architectures: a 325~million parameter transformer with no prior training, and a 12~billion parameter language model pretrained on a large corpus of high-quality data spanning multiple domains and comprising a mix of curated and synthetically-generated material\footnote{See  \cite{nvidia2025nvidianemotronnano2}, section 2.2 for full dataset details.}. The smaller model is first pretrained on a next-token prediction objective to reproduce reference circuits generated by ADAPT-VQE, and is then iteratively improved beyond ADAPT-VQE through a post-training scheme that combines RL and self-distillation.
The larger model is specialized to the circuit generation task via continual pretraining~\cite{ke2023continualpretraininglanguagemodels} and supervised fine-tuning. Figures~\ref{fig:gemma_architecture} and \ref{fig:nemo_architecture} provide a visual summary comparing the two model architectures and their respective training pipelines.

Throughout this section, we use the term \textit{pretraining} to refer to training with teacher forcing on ADAPT-VQE reference data, including both the continual pretraining and supervised fine-tuning stages of the larger model. We use \textit{post-training} to refer to the subsequent iterative improvement of the model using only model-generated circuits, through RL and self-distillation. We emphasize that the fine-tuning of the large model is supervised using ADAPT-VQE-generated circuits as a target and is therefore included under the umbrella of pretraining. In contrast, post-training of the small model is driven entirely by rewards assigned to model-generated circuits, without reference to ADAPT-VQE sequences.

Independent models are trained for each qubit count. This is necessitated by the fact that operators are represented by their indices within a specified operator pool, which in turn demands a fixed operator pool and active space across a given dataset. Extending this framework to enable generalization across different active spaces would require a pool-independent operator representation and a more general Hamiltonian encoding, which we leave for future work.




\subsubsection{Operator Language and Tokenization}
\label{subsubsec:tokenization}

For model training and inference we represent quantum circuits as strings of \opcoeffpair pairs, $\sum_k \mathcal{O}_k c_k$. Here $\mathcal{O}_k$ is an operator chosen from the operator pool and the sum indicates string concatenation. Coefficients $c_k$ are rounded to 6 decimal places to balance accuracy with sequence length and occur after operators in the sequence, allowing them to be conditioned on their respective operators during generation. 

For the smaller Gemma~3 model with no prior pretraining, we encode sequences using an untrained word-level tokenization: each operator $\mathcal{O}_i$ in the operator pool is assigned a unique token, and coefficients are represented with per-digit tokenization where each numerical character $({0 ... 9}, ., -)$ is a single token. The remainder of the vocabulary comprises start, stop, unknown and padding tokens, in addition to a token for injecting molecular information as multimodal input.

The larger pretrained Nemotron~Nano~2 model uses its original byte-pair encoding (BPE) tokenizer (vocabulary size $131,072$) inherited from the base model, with the additional multimodal Hamiltonian token added to the vocabulary. An operator at index $i$ in the operator pool is represented as the literal string $i$; thus in contrast with the smaller model, where each operator index receives its own token, the BPE may split a string for an operator token into up to 3 individual tokens. 

\subsubsection{Multimodal Hamiltonian Representation and Encoding}
\label{subsubsec:multimodality}

To condition circuit generation on information about the molecular configuration of a particular conformational state, we treat the molecular Hamiltonian as multimodal input: for both architectures, this is processed by a dedicated encoder module whose output is injected at the embedding layer of the language processing module. This multimodal approach allows complete information about the Hamiltonian to be provided as model input without enumerating the terms and coefficients as text input, which would require prohibitively long context lengths. While we explored using geometrical information from MACE \cite{mace-off} as multimodal input, we found providing the encoder with a vector of canonically-ordered Hamiltonian coefficients to be more effective for our task (see Appendix~\ref{appendix_sub:molecular_representation}), so we use that representation for results in the main text.

The vector representation is constructed as follows: each Hamiltonian is expressed in the qubit representation as a weighted sum of Pauli strings, $H = \sum_{i=1}^{d} h_i P_i$, where $P_i \in \left(I, X, Y, Z\right)^{\otimes n}$ are $n$-qubit Pauli operators, $h_i$ are real-valued coefficients, and $d$ is the number of Pauli string terms in the Hamiltonian. In our case, the $P_i$ are identical across conformers for a given number of active orbitals and electron count. By ordering the $P_i$ in a canonical way, this permits a compact representation of the Hamiltonian via a vector of the coefficients $h_i$. The Hamiltonian coefficients $h_i$ are normalized to the range $[-1, 1]$ by scaling by the maximum absolute value of each coefficient across the training data. We note that in general, the number of terms $d$ in the molecular Hamiltonian scales as $O(N^4)$ when using second quantization; if we were to enumerate the coefficients as text input, this would lead to an explosion in the required context length as qubit count increases.

To map the vector representation to the latent space of the text modules of either architecture, we use a learned encoder $f_\text{enc}: \mathbb{R}^{d} \to \mathbb{R}^{D}$ comprising an input projection, fully-connected residual blocks with layer normalization and SiLU activations, and a final projection to the text module embedding dimension. The output of the Hamiltonian encoder then replaces the embedding of the multimodal Hamiltonian token from section~\ref{subsubsec:tokenization}.

For the smaller model we place the multimodal token after each \opcoeffpair pair throughout each sequence so that for $k$ operators, the transformer attention mechanism can attend to the same molecular information from $k$ different positions interspersed throughout the context window. While deterministic placement of this token is only possible for the ADAPT-VQE dataset, the causal language modelling loss trains the model to insert this token in the same positions during generation. This token placement is atypical for vision-language models, where multimodal tokens only occur in a single contiguous block per-image; our choice is informed empirically after observing performance improvements compared with injecting a single token at the start of the sequence. For the larger pretrained model, a multimodal Hamiltonian token is placed once at the end of the natural language input prompt (Appendix~\ref{appendix:nemotron_architecture}).

Further details on the normalization procedure, encoder architecture, and hyperparameters can be found in the Appendix~\ref{appendix:encoder}. 

\begin{figure}[t]
    \centering
    \includegraphics[width=0.9\linewidth]{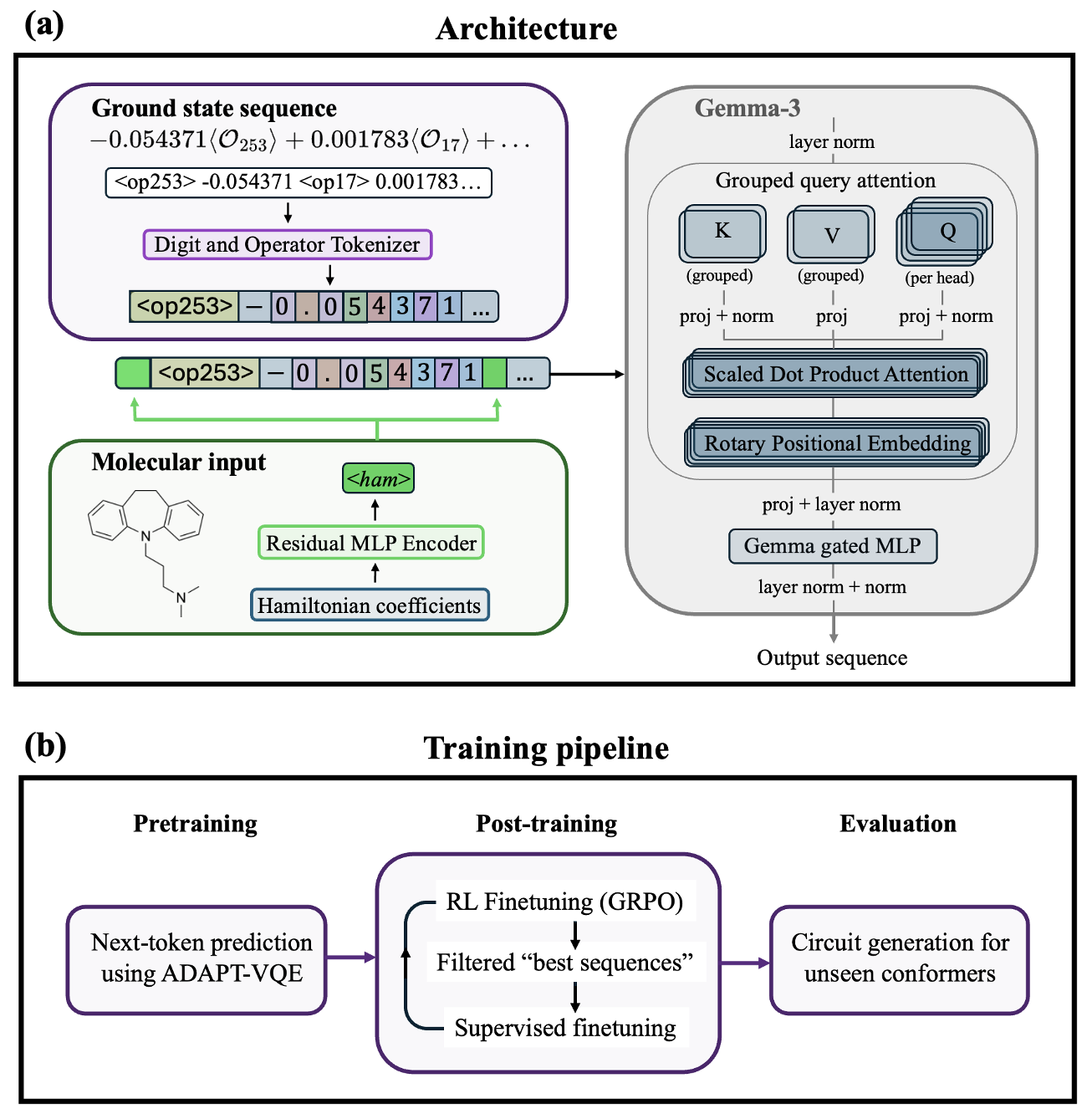}
    \caption{\textbf{Trained-from-scratch transformer architecture and training.} \textbf{(a)} ADAPT-VQE ground-state operator sequences are tokenized, with operators and numerical digits represented as individual tokens. Pauli-term coefficients from the molecular Hamiltonian are encoded with a residual MLP and interleaved within the token sequence before processing by the Gemma 3 transformer. \textbf{(b)} During pretraining, the model is trained for next-token prediction on ADAPT-VQE-generated sequences. During post-training, the model is iteratively refined through RL and supervised fine-tuning on model-generated sequences.
    }
    \label{fig:gemma_architecture}
\end{figure}

\begin{figure}[t]
    \centering
    \includegraphics[width=0.9\linewidth]{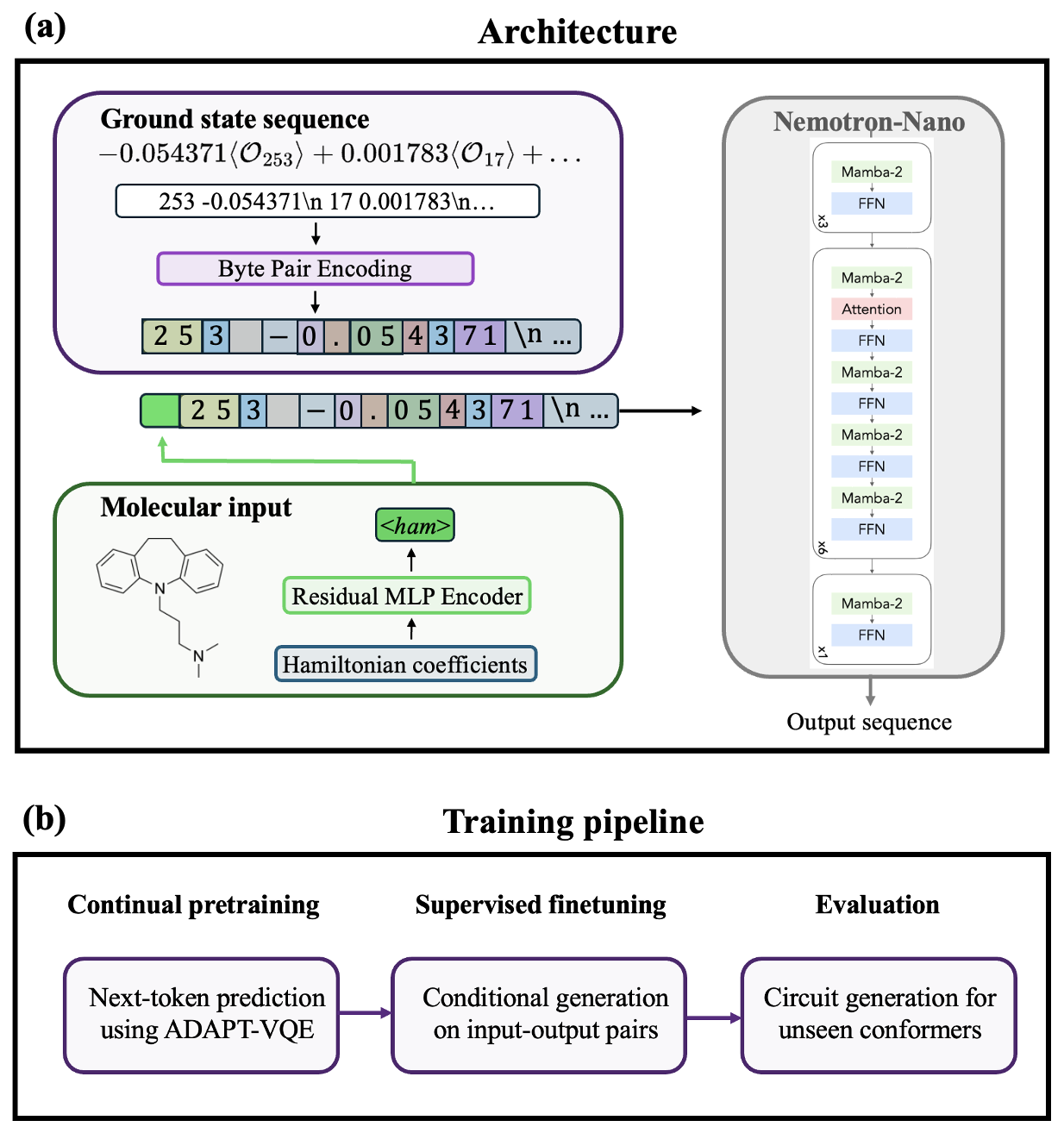}
    \caption{\textbf{Pretrained NVIDIA Nemotron architecture and training.} \textbf{(a)} ADAPT-VQE ground-state operator sequences are represented by their corresponding operator-pool indices and coefficients, and tokenized using the standard BPE tokenizer. The coefficients of the Pauli terms in the molecular Hamiltonian are encoded with a residual MLP and prepended to the token sequence before being processed by the Nemotron Nano transformer. \textbf{(b)} The pretrained model is first adapted to the operator language through continual pretraining and then specialized for circuit generation via supervised fine-tuning on ADAPT-VQE-generated sequences. 
    }
    \label{fig:nemo_architecture}
\end{figure}

\subsubsection{Gemma~3 Architecture and Training}
\label{subsubsec:gemma3-training}

We use the Gemma~3 architecture as a backbone for the small trained-from-scratch model. Gemma~3 serves as a representative example of a modern decoder-only language model using optimization techniques including grouped-query attention (GQA)~\cite{ainslie2023gqatraininggeneralizedmultiquery}, sliding window attention~\cite{beltagy2020longformerlongdocumenttransformer}, pre/post-normalization with RMSNorm~\cite{zhang2019root}, and rotary positional embeddings (RoPE)~\cite{su2024roformer}. The full model comprises the Gemma~3 text-only transformer architecture (no vision encoder) paired with the Hamiltonian encoder described in Section~\ref{subsubsec:multimodality}, and training of causal language modelling on the ADAPT-VQE circuits followed by a post-training phase involving iterations of a RL and self-distillation loop. A summary of the architecture and training of this model is shown in Figure~\ref{fig:gemma_architecture}.

\textbf{Pretraining:} The model is pretrained on causal language modelling over the ADAPT-VQE circuits, using the standard negative log likelihood loss of predicting the next token in an ADAPT-VQE sequence conditioned on the corresponding Hamiltonian coefficients. Training is terminated using an early-stopping criterion based on validation loss, and the checkpoint with the lowest validation loss is retained.

\textbf{Reinforcement learning:} Following pretraining, we further optimize the model using RL, where generated circuits are scored by their final energy rather than by token-level agreement with ADAPT-VQE. We use Group Relative Policy Optimization (GRPO)~\cite{shao2024deepseekmathpushinglimitsmathematical} with the DAPO variant of the GRPO loss~\cite{yu2025dapoopensourcellmreinforcement}. We use DAPO to reduce length-dependent bias in the reward assignment via token-level normalization, and use a formulation that accounts for multiple gradient updates per generation. Given the sparse nature of good candidates in the space of possible circuits the model can generate, we include a KL-divergence penalty to keep  changes to the policy from any single update conservative.

The reward function we use for RL is 
\begin{equation}
    R = \min\left(\frac{E_{\text{HF}} - E_{\text{ref}}}{E_\text{circuit} - E_{\text{ref}}} - 1,\ R_\text{max}\right)
    \label{eq:casci_reward}
\end{equation}
where $E_{\text{HF}}$ and $E_{\text{ref}}$ are the Hartree-Fock and reference energies associated with the conformational state of imipramine, respectively. For 12- and 14-qubit systems $E_\text{ref}$ is the CASCI energy and for the 16-qubit system it is the CCSD energy. $E_\text{circuit}$ is the energy of the valid part of the generated circuit\footnote{The model is capable of generating invalid sequences, for example by producing coefficients that are not numbers (e.g., with several decimal points). Invalid parts of generated sequences are discarded during training. For trained models, the mean valid fraction of generated sequences exceeds $0.9995$, and fewer than 1\% of sequences contain any parsing error.}. To improve stability, we clip the reward at a maximum value of 30 \--- a value which typically only occurs for circuits within chemical accuracy of $E_{\text{CASCI}}$. 
This mitigates the destabilizing effect of sampling rare extremely low-energy circuits, which would otherwise have an unbounded reward. 

One caveat of this particular reward function is that it requires a reference energy such as the CASCI energy, which may be expensive to compute for molecular systems with large active spaces. In Appendix~\ref{appendix_sub:reward_function} we show that comparable, albeit slightly lower, performance can also be achieved with an alternative reward function that does not require a reference energy.

\textbf{Self-distillation:} During GRPO, we log all sequences sampled by the model with their energies to a database seeded with the original ADAPT-VQE circuits. After each GRPO run, we filter this database to retain the top-$k$ lowest-energy sequences per-conformer below a specified energy threshold, and use these as a supervised fine-tuning (SFT) dataset. This procedure mirrors distillation approaches used in large language models~\cite{deepseekai2026deepseekr1incentivizingreasoningcapability}, and because the model generates the data used for its own fine-tuning, we refer to this process as self-distillation~\cite{yang2024selfdistillationbridgesdistributiongap}.

The RL and self-distillation procedure is repeated over multiple rounds with progressively stricter selection criteria (smaller $k$ and tighter energy thresholds). Further details on the training procedure and specific hyperparameters for the training can be found in Appendix~\ref{appendix_sub:gemma_parameters}.

\subsubsection{Nemotron~Nano~2 Architecture and Training}
\label{subsubsec:nemotron-training}

As an example of using a pretrained language model in our framework, we train NVIDIA Nemotron-Nano-12B-v2-Base~\cite{nvidia2025nvidianemotronnano2}. Nemotron~Nano~2 serves as a representative example of a language model pretrained on a large corpus of high-quality data spanning multiple domains and comprising a mix of curated and synthetically-generated material. Further, the model is an example of a hybrid Mamba--Transformer architecture, comprising Mamba-2~\cite{dao2024transformersssmsgeneralizedmodels}, self-attention, and feedforward layers. This design allows high throughput for reasoning workloads, interleaving efficient linear-time sequence processing from Mamba-2 blocks with full self-attention layers for long-range token interactions. The architecture contains 28 Mamba-2 layers, 28 feedforward layers, and 6 self-attention layers; see section 2.2 in \cite{nvidia2025nvidianemotronnano2} for details. A summary of the architecture and training of this model is shown in Figure~\ref{fig:nemo_architecture}.

 We train the model using continual pretraining (CPT)~\cite{ke2023continualpretraininglanguagemodels} followed by supervised fine-tuning (SFT). The two-stage design aims to decouple domain adaptation, where the model learns the concepts and language of operator sequences for state preparation, from task specialization, where it learns to effectively produce low-energy circuits. This choice is motivated by the observation that jointly learning Hamiltonian representations and instruction-following from a single objective leads to suboptimal encoder convergence. By first establishing a robust mapping from Hamiltonian structure to language model representations during CPT, then sharpening circuit generation behavior via SFT, we achieve better final accuracies than using either stage in isolation.

\textbf{Continual Pretraining:} We use CPT to train the custom Hamiltonian encoder and adapt the Nemotron-Nano-12B-v2-Base weights to the new Hamiltonian multimodality and our domain-specific syntax for operator sequences. During this stage, we train a randomly initialized Hamiltonian encoder in tandem with the pretrained Nemotron~Nano~2 using the causal language modelling objective on sequences formatted with a simple custom chat template (see Appendix \ref{appendix_sub:nemotron_parameters}). The loss is computed over all tokens: this includes the system message, user prompt, and assistant response.

\textbf{Supervised Fine-tuning:} SFT is used to specialize the model to precise circuit generation, conditioned on a Hamiltonian. In contrast to CPT, all prompt tokens (the system message, user prompt, and multimodal Hamiltonian token) are masked and excluded from the causal language modelling objective. This focuses learning entirely on producing correctly formatted circuit outputs.

More details of the architecture and training are in Appendix~\ref{appendix_sub:nemotron_parameters}.
Table~\ref{tab:cpt_hyperparams_Nemotron} and  Table~\ref{tab:sft_hyperparams_Nemotron} in Appendix~\ref{appendix:nemotron_architecture} specifically summarize the CPT and SFT hyperparameters employed for training. 

\section{Results}
\label{sec:results}

In this section we evaluate the effectiveness of ADAPT-GQE for ground state preparation along three axes: circuit generation accuracy, computational advantage in terms of time and scaling relative to ADAPT-VQE, and feasibility of circuit execution on quantum hardware. Section~\ref{sec:model_accuracy} assesses the ability of both model architectures --- Gemma (trained from scratch) and Nemotron (pretrained LLM) --- to generate accurate ground-state circuits for unseen conformers, characterizing performance as a function of system size, circuit complexity, and amount of training data. Section~\ref{subsec:speedup} quantifies the computational advantages of ADAPT-GQE relative to ADAPT-VQE by replacing iterative variational optimization with a model-generated circuit. Section~\ref{subsec:hardware_results} demonstrates that model-generated circuits are directly executable on state-of-the-art quantum hardware by reporting energy evaluations from the Quantinuum Helios trapped-ion processor. 

To quantify the accuracy of our generative models, each dataset is split into 80\% training, 18\% validation, and 2\% test sets. Models are trained on the training set and evaluated on the validation set for hyperparameter tuning, and final performance is reported on the test set, which consists entirely of conformers not seen during training or validation. For results in this section, datasets are split randomly into training, validation and test sets to capture performance on the dataset distribution, but we also show comparable performance for model generalization to out-of-distribution test conformers in Appendix~\ref{appendix_sub:generalization}. For all results, each model generates 16 candidate circuits for each conformer in the test set, and the circuit with the lowest energy for each test-set conformer is used in the reported metrics and figures. Details on the sampling hyperparameters for Gemma and Nemotron models are given in Appendices~\ref{appendix_sub:gemma_parameters} and \ref{appendix_sub:nemotron_parameters}. Corresponding numerical data for all results shown in this section are tabulated in Appendix~\ref{appendix:tab_results}.

Throughout this section, results are compared against two primary references. The imitation accuracy during pretraining is evaluated relative to the energies of ADAPT-VQE sequences that comprise the training data. However, because the ultimate objective is accurate ground-state preparation, the more meaningful performance metric is the energy difference between the generated circuits and the exact ground state solution within the active space. Throughout this Section, we use the mean $\pm$ standard deviation to characterize statistical distributions.

\subsection{Accuracy of Model-Generated Circuits}
\label{sec:model_accuracy}

\begin{figure}[htbp]
    \centering
    \includegraphics[width=\textwidth]{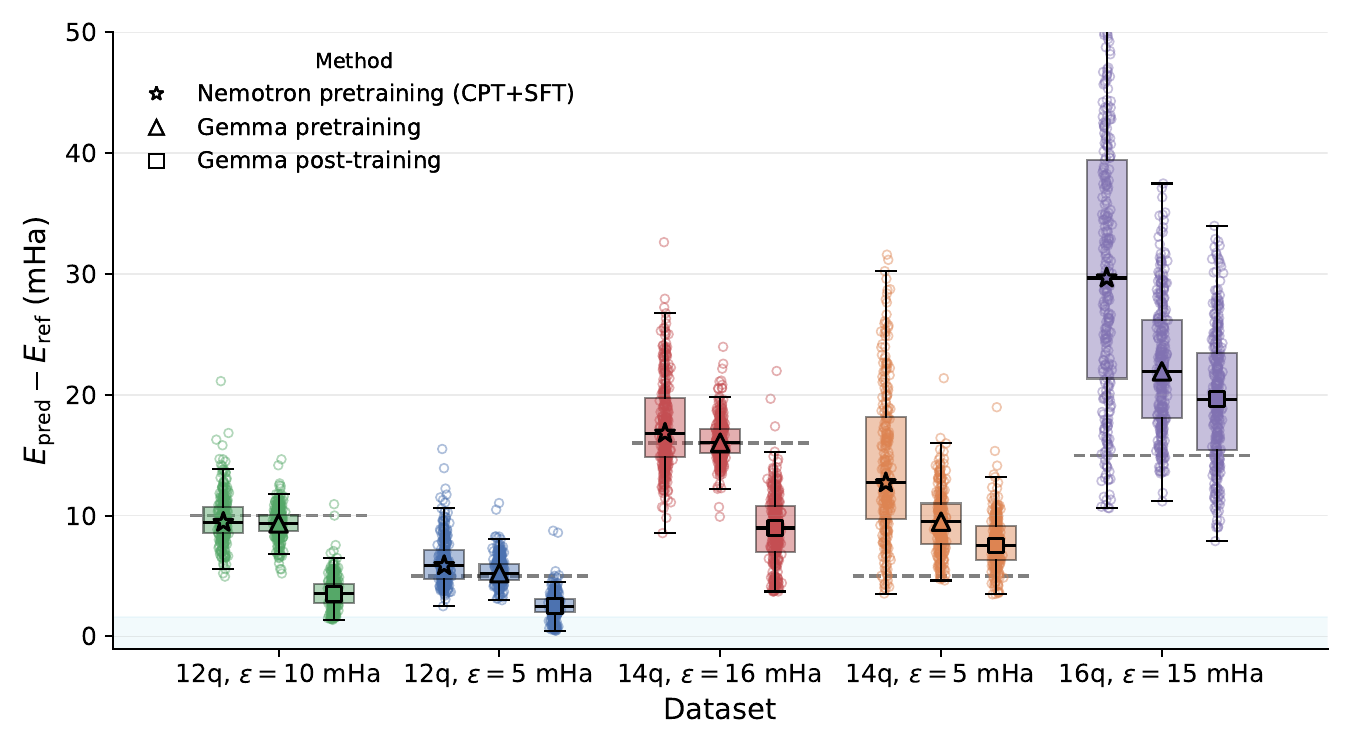}
    \caption{\textbf{Generated circuit energy errors for Nemotron and Gemma models.} Generated circuit energy errors after pretraining on ADAPT-VQE data (stars and triangles, respectively) and from the Gemma model after post-training (squares) for all datasets (indicated by color). For each case, individual data points are shown as circles, with summary statistics overlaid as a box-and-whisker plot (the central mark indicates the median, the box spans the interquartile range, and the whiskers extend over the non-outlier range). Energy tolerances $\varepsilon$ of the ADAPT-VQE training datasets are shown for reference with grey dashed lines. Models after pretraining faithfully mimic the accuracy of both 12-qubit datasets and the higher tolerance 14-qubit dataset, but degrade in accuracy for the more complex 14-qubit low tolerance and 16-qubit datasets. Post-training the small model improves over the pretraining accuracy in all cases. }
    \label{fig:mean-energy}
\end{figure}

Figure \ref{fig:mean-energy} shows the energy difference between the best of 16 model-generated circuits for each test conformer and the reference ground state energy for the model architectures and datasets described in the previous sections. 

\subsubsection{Pretraining on ADAPT-VQE data}

We begin by considering the pretraining results for both architectures in Fig.~\ref{fig:mean-energy}. During pretraining, the models are trained to reproduce the ADAPT-VQE operator sequences and associated coefficients (rotation angles). Because circuits in the training data are generated using an ADAPT-VQE convergence threshold $\varepsilon$, they satisfy $E - E_{\mathrm{CASCI}} \leq \varepsilon$ for the 12- and 14-qubit systems and $E - E_{\mathrm{CCSD}} \leq \varepsilon$ for the 16-qubit system. The threshold $\varepsilon$ therefore provides a natural benchmark for generated circuits: generated circuits with an energy error near or below $\varepsilon$ indicate successful reproduction of the training-set quality, while substantially larger errors indicate degradation relative to the ADAPT-VQE reference. To evaluate the performance of models trained to imitate ADAPT-VQE data we use $1$~mHa as a target accuracy (relative to ADAPT-VQE). This value is inspired by chemical accuracy ($\sim 1.6$~mHa) but is relative to the training set accuracy $\varepsilon$ rather than the true ground state energy, to quantify the ability of the model to imitate its training set. We use this value to present the percentage of generated circuits with energy between $1$~mHa above $\varepsilon$ and the CASCI energy.

For both 12-qubit datasets and the higher tolerance 14-qubit dataset, both model architectures are able to generate quantum circuits that mimic the energy accuracy of the ADAPT-VQE dataset. The error of the generated circuit energy from the convergence threshold of ADAPT-VQE is $0.77 \pm 3.71$~mHa (Gemma) and $1.37 \pm 1.89$~mHa (Nemotron) for the 12-qubit dataset with $\varepsilon=5$ mHa, and $-0.13 \pm 1.09$~mHa (Gemma) and $0.19 \pm 1.83$~mHa (Nemotron) for the 12-qubit dataset with $\varepsilon=10$ mHa. For the 14-qubit dataset with $\varepsilon=16$~mHa, generated circuits differ from the ADAPT-VQE threshold energy by $0.62 \pm 1.62$~mHa (Gemma) and $1.85 \pm 3.51$~mHa (Nemotron). We find that these represent good agreement with the ADAPT-VQE-generated datasets, as model-generated circuits typically fall within our target accuracy: $68.6\%$ (Gemma) and $49.5\%$ (Nemotron) for 12-qubit $\varepsilon=5$~mHa, $87.4\%$ (Gemma) and $73.8\%$ (Nemotron) for 12-qubit $\varepsilon=10$~mHa, and $67.6\%$ (Gemma) and $48.5\%$ (Nemotron) for 14-qubit with $\varepsilon=16$~mHa.

For the lower tolerance 14-qubit and 16-qubit datasets, we find that neither architecture is fully able to mimic the training set distribution after pretraining alone. The error of the generated circuit energy from the tolerance of the training dataset is $4.59 \pm 2.58$ mHa (Gemma) and $9.34 \pm 6.17$ mHa (Nemotron) for the 14-qubit dataset with $\varepsilon=5$ mHa, and $7.56 \pm 5.31$ mHa (Gemma) and $15.78 \pm 11.36$ mHa (Nemotron) for the 16-qubit dataset with $\varepsilon=15$~mHa\footnote{The interpretation of the 16-qubit results requires some care since the reference energy and ADAPT-VQE convergence threshold are defined relative to CCSD rather than CASCI.}. We attribute the degradation in pretraining accuracy primarily to the increased challenge of the language modelling task for the longer sequences in these datasets, which we will quantify in Section~\ref{subsec:error-complexity}. 
Despite the increased challenge, the models still produce physically meaningful circuits that recover substantial correlation energy beyond Hartree–Fock, demonstrating that the approach generalizes to larger qubit counts, albeit with reduced precision for more complex problems.

We generally find that the trained-from-scratch model produces a narrower distribution of generated energies, consistent with its specialization to the target problem during training. This concentration increases the likelihood of generating high-quality circuits. In contrast, the pretrained model exhibits a broader distribution of candidate solutions, reflecting the greater diversity retained from large-scale pretraining. Although this leads to larger variance in circuit quality, the increased diversity could provide advantages for exploration and subsequent refinement or distillation.

\subsubsection{Post-training with reinforcement learning and distillation}

We also show in Figure~\ref{fig:mean-energy} the results of performing post-training, where RL and distillation are used to fine-tune the trained-from-scratch Gemma model. We find that post-training in all cases is able to extend the model performance beyond the accuracy from pretraining. For the 12-qubit $\varepsilon=5$~mHa and  $\varepsilon=10$~mHa datasets, we achieve an improvement in the median energy over pretraining of $3.09$~mHa and $5.79$~mHa, respectively. For the 14-qubit $\varepsilon=5$~mHa and  $\varepsilon=16$~mHa datasets, we achieve an improvement of $1.73$~mHa and $7.31$~mHa, respectively. For the 16-qubit $\varepsilon=15$~mHa dataset the improvement is $1.99$~mHa. These improvements result in median differences to the CASCI energy of $2.53 \pm 1.08$~mHa for the 12-qubit $\varepsilon=5$~mHa dataset, $3.61 \pm 1.30$~mHa for the 12-qubit $\varepsilon=10$~mHa dataset, $7.78 \pm 2.17$~mHa for the 14-qubit $\varepsilon=5$~mHa dataset, $8.91 \pm 2.88$~mHa for the 14-qubit $\varepsilon=16$~mHa dataset, and $20.4 \pm 5.42$~mHa for the 16-qubit $\varepsilon=15$~mHa dataset.

For the 12- and 14-qubit systems, where datasets are available at two convergence tolerances, we observe substantially larger improvements relative to the training data for the looser-tolerance datasets for the same amount of post-training fine-tuning. This suggests that circuit-generation difficulty is governed more by how closely the target approaches the CASCI energy than by the absolute size of the improvement, presumably because higher-accuracy targets typically require longer and more complex circuits. More practically, these results indicate that post-training can effectively compensate for lower-accuracy ADAPT-VQE training data.

We performed post-training fine-tuning using the trained-from-scratch Gemma model rather than Nemotron Nano because of its improved training stability. The custom tokenizer used by the Gemma model represents equivalent circuits with substantially fewer tokens than the BPE tokenizer, reducing sequence lengths and inference costs. Furthermore, Nemotron Nano contains nearly an order of magnitude more parameters, increasing the number of samples needed for effective RL optimization. We also leveraged the multiple gradient updates per generation batch provided by the Hugging Face Transformers Reinforcement Learning (TRL) library~\cite{vonwerra2020trl}, which further improved sample efficiency. Collectively, these factors mitigated the length collapse and reward-hacking behaviors observed when applying GRPO to the Nemotron model.

\subsubsection{Correlation of Circuit Generation Error with Complexity}
\label{subsec:error-complexity}

\begin{figure}[htbp]
    \centering
    \includegraphics[width=\textwidth]{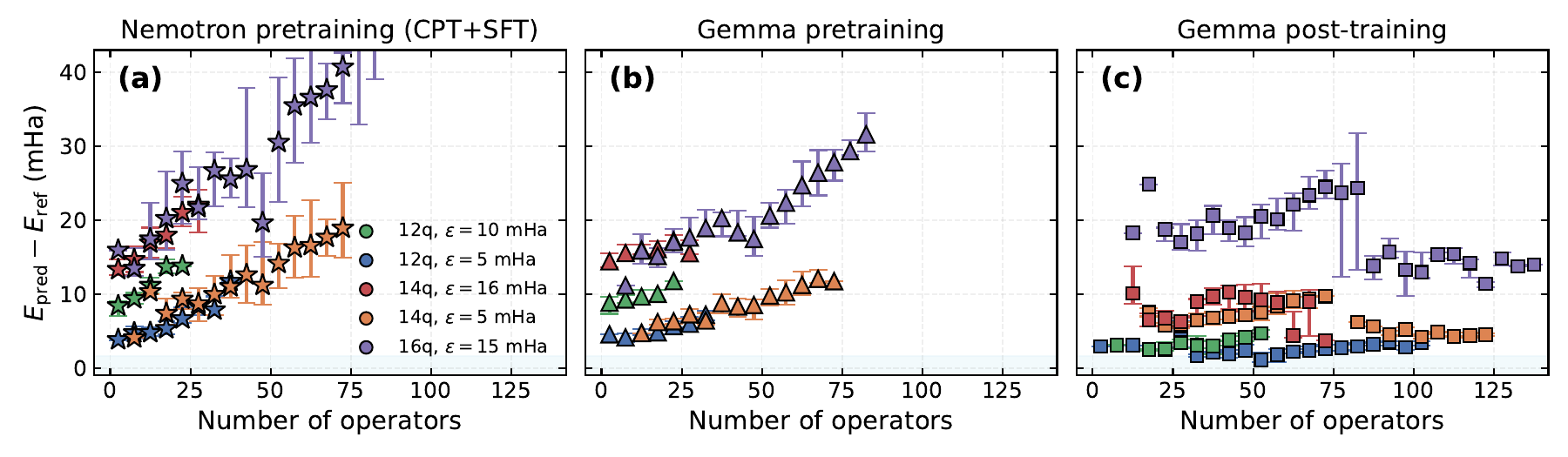}
\caption{\textbf{Dependence of energy errors on generated circuit complexity.} Generated circuit energy errors as a function of the number of operators in the generated sequence for \textbf{(a)} the Nemotron model after pretraining, \textbf{(b)} the Gemma model after pretraining, and \textbf{(c)} the Gemma model after post-training fine-tuning. Points are the distribution median, and error bars indicate the interquartile range. After pretraining, both model architectures show a strong correlation between the energy error of the generated circuit and the number of operators selected during generation. This correlation is largely removed after post-training of the Gemma model.}
    \label{fig:error-vs-ops}
\end{figure}

Across all datasets and both architectures, models trained by next-token prediction on ADAPT-VQE data exhibit a strong positive correlation between circuit complexity and the energy error of the generated circuits. Figure~\ref{fig:error-vs-ops} shows the energy prediction errors as a function of the number of operators in the generated sequence, a proxy for the circuit complexity. 
The Pearson correlation coefficients between the energy difference to the reference energy and the number of generated operators for all datasets are $0.03$--$0.72$ (mean $0.41$) for the pretrained Gemma model and $0.39$--$0.73$ (mean $0.59$) for the pretrained Nemotron model (see Table~\ref{tab:pearson_correlations} for complete numerical values). Conformers whose ADAPT-VQE reference circuits contain fewer than $\sim10$ operators are reproduced with near-zero or negative mean error relative to ADAPT-VQE (i.e., the model matches or improves upon ADAPT-VQE) for all datasets, whereas conformers requiring 20 operators exhibit mean errors of $2$--$6$~mHa with respect to ADAPT-VQE. 
This monotonic increase in error with circuit depth suggests that long-sequence generation is a primary bottleneck to obtaining good model accuracies with pretraining.

This effect explains the substantially larger median error of models after pretraining on the low tolerance 14-qubit and 16-qubit datasets shown in Figure~\ref{fig:mean-energy}.
As was shown in Fig.~\ref{fig:circuit-depth}, the typical ADAPT-VQE operator sequence lengths vary substantially across datasets. The low- and high-tolerance 12-qubit datasets and the high-tolerance 14-qubit dataset contain relatively short circuits, with median operator counts of $8$, $19$, and $15$, respectively, whereas the 14-qubit $\varepsilon=5$~mHa and 16-qubit $\varepsilon=15$~mHa datasets exhibit substantially deeper circuits with a median of 51 and 50 operators respectively. In all datasets, operator sequence lengths are concentrated near the median depth, with comparatively few very short or very long circuits. Despite this, the accuracy with which the models reproduce ADAPT-VQE circuits during pretraining degrades steadily with increasing operator count. This trend likely reflects a combination of autoregressive error accumulation during long-sequence generation and the greater structural complexity of conformers requiring deeper ADAPT-VQE circuits to achieve the same energy tolerance. This is consistent with the observation that Nemotron has a typically-higher correlation of error with depth, due to a larger error accumulation because the same circuit is represented with a larger number of tokens.

Fine-tuning the Gemma model on model-generated circuits during post-training substantially reduces its error on long sequences. This is shown in Figure~\ref{fig:error-vs-ops}(c), where the error in the energy of generated circuits improves substantially on the pretraining results for long circuits, and even renders the long circuits to be more accurate than shorter circuits in most cases. The Pearson correlation coefficients between the energy difference to the reference energy and the number of operators in this case reduce to the range $-0.2$--$0.32$ (mean $0.02$), with all three beyond 12 qubits being negative, indicating a much weaker and often negative correlation between error and the number of operators. After fine-tuning and distillation, the model accuracy on very long sequences is often even better than the accuracy on shorter sequences. This suggests that, by optimizing directly on producing low energies rather than focusing on imitating ADAPT-VQE, the model is able to 
effectively generate long circuits with much less per-token quality degradation than from pretraining.

Sequences generated by the model after fine-tuning tend to have many more operators than the ADAPT-VQE training data, with an increase in the median number of operators of $+50$ and $+32$ for the 12-qubit datasets with $\varepsilon=5$ and $\varepsilon=10$, respectively, and $+2$ and $+35$ for the 14-qubit datasets with $\varepsilon=5$ and $\varepsilon=16$, and $+8$ for the 16-qubit dataset (see Appendix~\ref{appendix_sub:ciruit_resource_requirements}). Interestingly, especially for the 12-qubit datasets, the model generates few sequences after fine-tuning with the intermediate operator numbers that were most common in the training data, and the improvement that we see in the model accuracy from fine-tuning compared to pretraining is also much more moderate in that regime. This highlights a key distinction between data generation for pretraining and RL: during pretraining, the dataset distribution is fixed by ADAPT-VQE, possibly with additional data targeted at long sequences, whereas in RL the model generates and scores its own training data, allowing the training distribution to shift more substantially in both circuit length and accuracy.

\subsubsection{Effect of Training Dataset Size}

To directly assess the importance of the size of the ADAPT-VQE dataset used for pretraining, we augment the training set with additional data for two configurations: the 12-qubit dataset at $\varepsilon = 5$~mHa was increased from 11,451 to 15,758 training samples (+37\%), and the 14-qubit dataset with $\varepsilon = 16$~mHa from 12,564 to 15,236 samples (+21\%). Test set sizes were increased proportionally (from 299 to 411 conformers for 12 qubit and from 330 to 401 conformers for 14 qubit).

\begin{figure}[htbp]
    \centering
    \includegraphics[width=\textwidth]{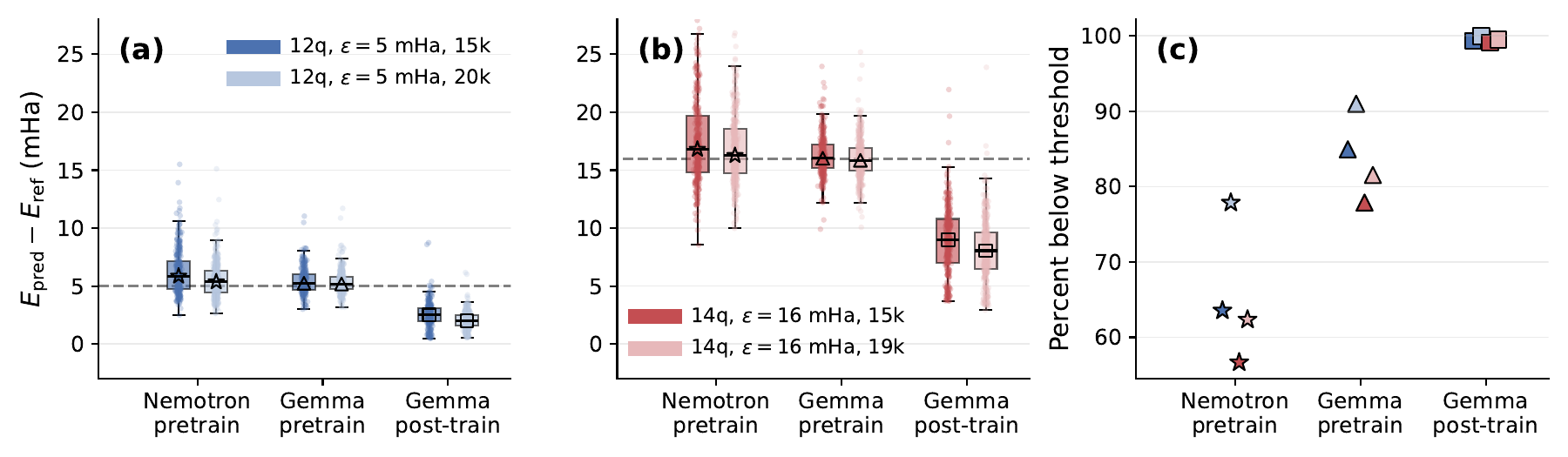}
    \caption{\textbf{Energy error comparison for different sizes of training dataset.} Comparison of generated circuit energy errors for Gemma and Nemotron models for active spaces with different sizes of ADAPT-VQE datasets for \textbf{(a)} 12-qubit $\varepsilon=5$~mHa and  \textbf{(b)} 14-qubit $\varepsilon=16$~mHa. The standard dataset sizes (shown in other figures) are dark colors and larger, augmented datasets are light colors. Larger training datasets moderately improve the median circuit generation accuracy and substantially decrease the spread in generated circuit quality, especially for Nemotron. \textbf{(c)} The resulting percentage of generated circuits with energies within the target accuracy of the ADAPT-VQE training data threshold, which improves with larger datasets for both models and saturates for all datasets after post-training fine-tuning.
    }
    \label{fig:data-scaling}
\end{figure}

Figure~\ref{fig:data-scaling} shows the effect of increased training data on both models. For both the 12- and 14-qubit datasets, the median energy error relative to the ADAPT-VQE tolerance decreases moderately with additional data, while the standard deviation decreases substantially, particularly for Nemotron. For the 12-qubit dataset the mean circuit energy reduces by $0.28$~mHa (Gemma) and $0.7$~mHa (Nemotron) and the standard deviation reduces by $2.77$~mHa (Gemma) and $0.44$~mHa (Nemotron). For the 14-qubit dataset the mean circuit energy reduces by $0.21$~mHa (Gemma) and $0.65$~mHa (Nemotron) and the standard deviation reduces by $-0.31$~mHa (Gemma) and $0.62$~mHa (Nemotron). The typical reduction in both the mean and variance of the gap to the reference energy indicates that larger datasets improve not only average accuracy but also the consistency of circuit generation across conformers. The stronger reduction in dispersion observed for Nemotron likely reflects its larger model parameter count and correspondingly greater sensitivity to data limitations.
Consistent with this trend, the fraction of high-quality circuits with energies within 1~mHa of the ADAPT-VQE threshold increases for both architectures. For the 12-qubit active space, this fraction increases from $68.6\%$ to $76.4\%$ for the Gemma model after pretraining and from $64\%$ to $78\%$ for the Nemotron model. For the 14-qubit active space, the corresponding improvements are $67.6\%$ to $69.3\%$ for Gemma and $55\%$ to $62\%$ for Nemotron.

These results suggest that further improvements in pretraining accuracy may still be achievable through larger-scale data collection, particularly with targeted sampling of high-operator-count conformers for the more difficult 14- and 16-qubit systems where quality of ADAPT-VQE circuits is not consistently achieved by model-generated circuits. However, increasing the size of the training dataset primarily reduces the variance of generated circuit quality, but cannot systematically improve performance beyond the convergence threshold defining the training distribution. Post-training fine-tuning, by contrast, substantially improves the median quality of generated circuits and the fraction of high-quality circuits saturates near 100\% across all datasets. While RL is less sample-efficient than pretraining, circuits are much more computationally efficient to generate than ADAPT-VQE data and RL allows the model to directly optimize the final energy objective instead of the per-token agreement with ADAPT-VQE-generated sequences.

\subsection{Computational Speedup: ADAPT-GQE vs. ADAPT-VQE}
\label{subsec:speedup}

In this section we quantify the computational benefit of using trained generative models for ground state circuit generation compared to using ADAPT-VQE. This impacts the utility of ADAPT-GQE as a potential replacement for ADAPT-VQE, and influences the effectiveness of training on model-generated data with RL.

The total inference cost for generative models includes circuit generation, in which the model produces 16 candidate circuits in a single batched call, and energy evaluation, in which all 16 candidates are executed as statevector simulations. For the Nemotron model, circuit generation is performed on a single GB200 node ($4\times$GB200, 186~GB each, NVLink 5th generation at 1.8~TB/s), and for Gemma it is performed on a single H100 GPU. Timings of energy evaluations for both models and of ADAPT-VQE are evaluated using CUDA-Q (FP64) on a single H100. To provide a controlled comparison, the timings reported here are measured for a single selected imipramine conformer at each qubit count.

The central computational bottleneck of ADAPT-VQE is its iterative structure: at every step, the energy gradient must be evaluated for every operator in the pool; the operator with the largest gradient is selected and appended to the ansatz; and all accumulated variational parameters in the expanded circuit are globally re-optimized. This process repeats until convergence, requiring a number of quantum circuit executions that grows as $O (\mathrm{pool\,size} \times d)$ where $d$ is the circuit depth. Repeated global re-optimization of all accumulated variational parameters adds a further optimizer-dependent cost, and the entire procedure must be repeated independently for every molecular geometry. The consequence is steep scaling: a 12-qubit system at $\varepsilon=5$~mHa requires 2,333~s of wall-clock time, which increases to 23,405~s for a 16-qubit system at $\varepsilon=15$~mHa — a $10\times$ increase for a mere 4-qubit enlargement of the active space, reflecting the exponential growth of the Hilbert space and the increasing pool size.

ADAPT-GQE eliminates this iterative loop entirely. Given a new molecular Hamiltonian, the fine-tuned language model generates a complete ground-state circuit in a single autoregressive forward pass — no gradient evaluation, no iterative operator selection, no per-geometry re-optimization. 
The generation stage accounts for over 99\% of total ADAPT-GQE wall time for the Nemotron model across all configurations: 164.0~s (12q), 293.3~s (14q), and 563.9~s (16q). For the much smaller Gemma model, inference is much faster --- 1.86~s (12q), 3.66~s (14q), and 1.62~s (16q). We attribute this improvement to three factors: Gemma has approximately $10\times$ fewer parameters than Nemotron, its custom tokenization produces shorter context lengths, and inference is accelerated using vLLM~\cite{kwon2023efficientmemorymanagementlarge}. Generation time increases moderately with system size for both the Nemotron and Gemma models. We summarize the wall-clock results and operator counts for both models across all three system sizes in Table~\ref{tab:model_speedup} in Appendix~\ref{appendix:compute}.

The end-to-end speedup relative to ADAPT-VQE is $14.1\times$ at 12 qubits and $41.4\times$ at 16 qubits for Nemotron, and $930\times$ to $8,800 \times$ for Gemma. This increasing advantage with system size arises because ADAPT-VQE scales steeply with the size of the active space — the $10\times$ increase in ADAPT-VQE cost from 12 to 16 qubits reflects the growing operator pool and the compounding cost of iterative gradient evaluation over more operators — while ADAPT-GQE generation time grows far more modestly, constrained primarily by output sequence length rather than Hilbert space dimension. Applied at the scale of hundreds of conformers per molecule, this translates to a reduction in total dataset generation time from weeks to minutes.

\subsection{Hardware Validation on Quantinuum Helios-1}
\label{subsec:hardware_results}

\begin{figure}[t]
    \centering
    \includegraphics[width=\textwidth]{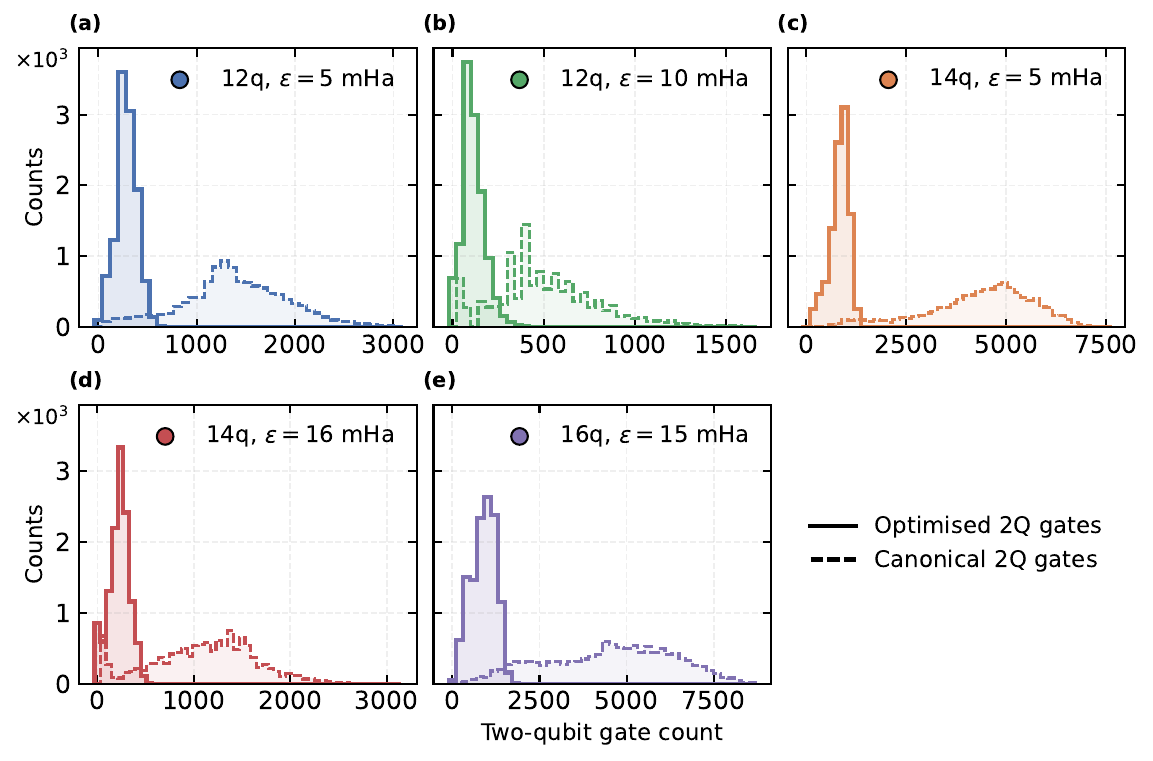}
    \caption{\textbf{Canonical and optimized two-qubit gate counts for each dataset.} Canonical (dashed) and optimized (solid) two-qubit gate counts for each ADAPT-VQE training dataset considered in this work. Panels \textbf{(a)} and \textbf{(b)} show the 12-qubit active space with $\varepsilon=5$~mHa and $\varepsilon=10$~mHa, \textbf{(c)} and \textbf{(d)} show the 14-qubit active space with $\varepsilon=5$~mHa and $\varepsilon=16$~mHa, and \textbf{(e)} shows the 16-qubit active space with $\varepsilon=15$~mHa.  Optimization dramatically reduces the required two-qubit gate counts relative to the canonical form for all datasets. Optimization also rebases the circuits to the native gateset of Quantinuum Helios-1.}
    \label{fig:2qgates}
\end{figure}

    So far, we have evaluated each generated circuit by its ideal circuit energy: the noiseless energy obtained from statevector simulation and compared directly with ADAPT-VQE and CASCI or CCSD energy benchmarks. This metric isolates the expressivity of the generated ansatz, but it does not capture whether the circuit can be executed reliably on current quantum hardware. We therefore next assess the resource requirements of the generated circuits relative to the capabilities of present-day devices.

    For circuits to be executed on a real device, fermionic excitations have to be converted into one and two-qubit gates.
    In the current noisy intermediate-scale quantum (NISQ)~\cite{Preskillnisq} era, noise accumulates in the device from all operations, with two-qubit gates being the most prominent source. In the canonical representation, fermionic excitation operators are converted into one qubit rotations entangled with CNOTs~\cite{yordanovencodefermionic}. Even the modest number of operators in Figure~\ref{fig:circuit-depth} can lead to thousands of two-qubit gates and extremely high depth in this canonical representation. Circuits of this scale, when executed without modification, would be overwhelmed by device noise. 

    To circumvent this, and make execution possible on current hardware, we leverage circuit optimisation techniques and device specific compilation for Quantinuum's Helios-1 trapped ion machine~\cite{ransford2025helios98qubittrappedionquantum} using \texttt{pytket}~\cite{pytket}. The compilation passes provided perform gate cancellation, commutation based reordering and gate synthesis~\cite{greedypaulisimp} and markedly reduce the two-qubit gate count from a few thousand, down to hundreds. Figure~\ref{fig:2qgates} shows the number of two-qubit gates for the datasets studied in this work, both in the canonical representation and after optimization. Optimization results in a factor $4$--$5$ improvement in the median number of gates for all datasets.

            \begin{table}
        \centering
        \begin{tabular}{cc|cccc}
             \hline \multicolumn{2}{c|}{Quantity}&CASCI&Statevector&Helios-1 Emulator & Helios-1  \\
             \hline\multirow{4}{*}{\makecell{Fine-tuned\\ Gemma}}&Energy (Ha)&-841.953249&-841.952101&-841.871250&-841.752841\\ 
             &Shot Noise (mHa)&-&-&11.7&18.1\\ 
             &$E-E_{\text{CASCI}}$ (mHa)&-&1.15&81.99&200.4\\ 
             &Discard Rate (\%)&-&-&22.76&48.74\\
             \hline\multirow{4}{*}{\makecell{Nemotron\\ SFT}}&Energy (Ha)&-841.920747 &-841.917254&-841.813407&-841.515128\\ 
             &Shot Noise (mHa)&-&-&11.9&20.6\\ 
             &$E-E_{\text{CASCI}}$ (mHa)&-&3.49&107.3&405.6\\ 
             &Discard Rate (\%)&-&-&20.73&33.28\\
             \hline
        \end{tabular}
        \caption{\textbf{Energy evaluation results for model generated circuits executed on Quantinuum Helios-1.} Energies are obtained via operator averaging with 500 shots. Error bars on Helios and emulator results reflect shot noise. The Gemma circuits were executed between 28 January and 8 Feb 2026; the Nemotron circuits were executed between 8 May and 18 May 2026.}
        \label{tab:hardware}
    \end{table}

    Even with optimization, device noise can still be prevalent and subsequent erroneous measurements can lead to systematic errors in the calculated energies. Rather than applying expensive error correction techniques (which add both a classical and large quantum resource overhead by significantly increasing the circuit resources and depth), we apply Partition Measurement Symmetry Verification (PMSV)~\cite{pmsv}. This exploits any symmetries present in the Hamiltonian, in this case particle number, by preparing additional Pauli strings (stabilisers) to be measured alongside the terms in the Hamiltonian. Shots whose measurement outcomes violate the expected symmetry are treated as invalid and discarded. This approach adds only minimal quantum overhead, requiring only a small number of additional single-qubit gates, while the symmetry check itself is performed entirely in classical post-processing and can improve the accuracy of the estimated result. However, the sampling overhead of symmetry post-selection grows exponentially with system size, limiting its scalability. As such, it is primarily useful for improving the performance of small-scale or shallow circuits.

    To demonstrate the whole framework end-to-end, we select one 12 qubit circuit generated by the Gemma model after RL fine-tuning, and one from the Nemotron model after supervised fine-tuning, both consisting of 18 operators. In the canonical representation, these contain 1440 and 1248 two-qubit gates which are optimised down to 244 and 283 two-qubit gates respectively. Circuit compilation, device submission, and energy evaluation are performed using Quantinuum’s InQuanto computational chemistry platform and the Nexus cloud access platform~\cite{inquanto,qnexus}. Energies are calculated with operator averaging with a mere 500 shots per circuit, a deliberately low number chosen to assess the hardware performance under near term practical resource constraints. Table \ref{tab:hardware} contains the results across three different evaluation modes: statevector simulation, Helios-1 emulator (fitted noise model), and Helios-1 (real device) alongside the corresponding CASCI energy. 

    For the Gemma circuit, the statevector energy lies within chemical accuracy of the CASCI energy with a difference of $1.15$~mHa. The Nemotron circuit lags only slightly behind, producing an energy $3.49$~mHa away. This confirms the capability of the models to produce highly accurate ground state circuits within the tolerance of the ADAPT-VQE data from the CASCI, as seen in Figure \ref{fig:mean-energy}. The corresponding emulator results for the models show a modest deviation away from the noiseless calculation, producing energy differences to the CASCI of $81.99$~mHa and $107.3$~mHa respectively. Despite the low shot count, the error bars on these calculations are just shy of $12$~mHa indicating that it is the noise model accounting for the majority of the difference rather than shot noise. Even with a minimal error mitigation technique in PMSV applied, with such a low shot count the accumulated gate errors shift the expectation value away from its ideal value in such a manner that the noiseless value cannot be fully recovered by symmetry post-selection alone.    
    
    The corresponding hardware results diverge significantly more from the reference, differing to the CASCI energy by $200.4$~mHa and $405.6$~mHa respectively, alongside a small increase in the shot noise, to $18.1$~mHa and $20.6$~mHa. The deviation between the hardware and emulator results can be rationalised by considering the proportion of shots discarded due to the application of PMSV. In both emulator cases, as seen in Table \ref{tab:hardware}, invalid symmetry shots corresponded to approximately 20\% of the total shots, whereas in the hardware runs, for the Gemma model circuit it was closer to 50\% and 35\% in the Nemotron model circuit. This suggests that not only was the emulator model not capturing all the behaviour of the device, but also that the hardware results may not be statistically significantly different from the emulator results due to the notably small shot count which could lead to undersampling of the expectation value measurements.

    While some of the deviation from hardware to emulator (and analogously to the statevector) results could be alleviated by taking more shots, or allocating them in an adaptive way by considering the norm of the term in the Hamiltonian or by the variance, these would still not recover the full accuracy needed for quantum chemistry calculations. In order to close the gap between statevector simulations and hardware evaluation, stronger error mitigation and correction techniques are necessary.

\section{Discussion}
\label{sec:discussion}

ADAPT-GQE reframes the role of variational quantum algorithms: rather than treating ADAPT-VQE as a solver, we use it as a method to generate high-quality training data and cast circuit generation as a structured sequence-learning problem. Pretraining models for next-token prediction with ADAPT-VQE data as a reference drives models to generate circuits that reproduce ADAPT-VQE quality by design. We find this to be highly effective for active spaces with relatively short ADAPT-VQE sequence lengths, but the replication ability degrades for cases where the ADAPT-VQE sequences are longer. The primary limitation of pretraining is performance degradation for sequences with a high number of operators: errors in next-token prediction accumulate over long sequences, with each error taking the generated circuit further from the training distribution. 
Through post-training with RL and knowledge distillation we drive the accuracy of model-generated sequences significantly beyond the ADAPT-VQE data used for training. Unlike models trained to directly reproduce ADAPT-VQE data, optimizing on the final objective (energy minimization) allows the model to maintain comparable or better performance for long sequences than for short ones. 

The eventual goal of this research direction is to provide a scalable approach for ground state circuit generation for large problems of industrial relevance where it may become computationally intractable to produce ADAPT-VQE data in large quantities. We have outlined a strategy to use RL fine-tuning and distillation to systematically improve the model and generate circuits that are better than anything seen in the training data by having the model propose and score its own circuits. In the future, an important milestone for scalability would be reduced reliance on the (expensive) training data. In the most extreme case, it may be possible that pretraining could be used primarily to learn syntax and generate typical circuits, with the bulk of the optimization for per-conformer circuit generation being handled by the RL.

To realize this transition, it will be critical to improve the efficacy and efficiency of learning to generate circuits with RL. An important challenge of RL in this work is the sparse reward -- the model only receives feedback on the quality of the solution after the whole circuit is generated, based on the energy of the whole circuit. Sparse rewards in RL pipelines are a well-known problem that can make it difficult for the model to assign credit for especially good (or bad) operator choices~\cite{Sutton1998,pignatelli2023survey}. We expect that assigning rewards for each operator choice could substantially improve training, with the disadvantage that doing this naively would introduce a substantial additional computational cost of evaluating the energy of many partial sequences. A possible solution would be to use a trained model to replace the sparse terminal reward with dense per-step supervision derived directly from ADAPT-VQE trajectories, as in process reward models ~\cite{Lightman2024VerifyStepByStep,Wang2023MathShepherd}.

A further challenge in difficult decision-making problems of this kind is that the utility of a given decision may depend strongly on the surrounding sequence of decisions. Explicitly incorporating longer-horizon planning into the training or generation procedure may therefore improve performance and enable the discovery of more optimal solutions. Monte Carlo tree search (MCTS) offers one promising approach for adding such structure to the search over operator sequences. By balancing exploration and exploitation, MCTS can guide search through complex, high-dimensional decision spaces. In a hybrid framework, MCTS could be used to select among promising operator choices, while the language model generates or optimizes the corresponding coefficient conditioned on that choice. We further anticipate that performance could benefit from more explicit modeling of coefficient sign and magnitude, rather than relying entirely on per-digit autoregressive generation.

The dramatic computational advantage of ADAPT-GQE makes the framework practical for ensemble-scale applications where ground-state energies must be evaluated across large numbers of problem instances. We find $1$k- to $8$k-fold speedups over ADAPT-VQE for the trained-from-scratch Gemma model and $14$- to $86$-fold speedups for the Nemotron model. This advantage becomes more pronounced with increasing system size, as the computational cost of ADAPT-VQE grows more rapidly than that of ADAPT-GQE for larger systems. This represents an enormous advantage of generative models over state-of-the-art computational techniques.

We have demonstrated that ground states generated by these models are executable on current state-of-the-art quantum hardware. The gap between noiseless simulation and real-device results reflects the importance of sophisticated error correction techniques on modern devices, and will narrow as error mitigation and early fault-tolerant techniques improve. Together this work provides a proof-of-concept demonstration of the power of generative AI for quantum chemistry with applications to near-term quantum hardware.

\section{Conclusion}
\label{sec:conclusion}

We have demonstrated that language models can generate quantum circuits for preparing the ground state of an industry-relevant molecular system, providing a first proof-of-concept of language-model-based ansatz generation for quantum chemistry. Our approach uses reinforcement learning to systematically improve beyond the initial training dataset, offering a potential route to reduce dependence on classically generated examples from methods such as ADAPT-VQE. Our results suggest that ground-state preparation circuits possess learnable structure that can be captured by sequence models and transferred across molecular geometries.


The approach offers a particularly important advantage in inference cost. In contrast to adaptive methods such as ADAPT-VQE, which construct circuits through repeated operator selection and optimization, the trained model amortizes this search into an offline training stage. Once trained, ADAPT-GQE generates candidate circuits in a single forward pass, without iterative optimization or quantum evaluations during circuit generation. This substantially reduces the computational cost of inference and yields more favorable scaling with system size, demonstrating the potential for scalable circuit synthesis.

We have further shown the first proof-of-concept that the resulting quantum circuits can be evaluated on current state-of-the-art quantum hardware. This hardware demonstration does not yet achieve sufficient accuracy for chemistry usage, but it shows that the generated circuits are compatible with execution on present-day devices and provides a concrete testbed for future improvements in hardware, error mitigation, and error correction.

A key direction for future work is to develop language models that can prepare ground states not only across different configurations of the same molecule, but also across distinct molecular systems. This remains a challenging problem, as it requires reformulating the language-modeling task using an operator language and molecular representation that encode chemical structure in a way that generalizes across active spaces and molecular species. If achieved, such a capability would have significant implications for quantum chemistry and would provide a compelling motivation for the co-development of generative AI and quantum computing.

\section{Acknowledgments}
 
This research used resources of the National Energy Research Scientific Computing Center, a DOE Office of Science User Facility supported by the Office of Science of the U.S. Department of Energy under Contract No. DE-AC02-05CH11231 using NERSC award NERSC DDR-ERCAP0033753 (E.~Rinaldi) and NERSC DDR-ERCAP0036478 (M.~Farag).
QNTM authors acknowledge valuable contributions from Eric~Brunner, Simon~McAdams, and Blake~Wilson.

\bibliography{machine-learning, computational-chemistry}

\appendix 
\section{Supplementary Material}
\addcontentsline{toc}{section}{Supplementary Material}

\subsection{Supporting Results}
\label{appendix:results}

In this section we provide a set of auxiliary results that support claims or conclusions provided in the main text. 

\subsubsection{Model generalization to out-of-distribution configurations}
\label{appendix_sub:generalization}

\begin{figure}[h!]
    \centering
    \includegraphics[width=\textwidth]{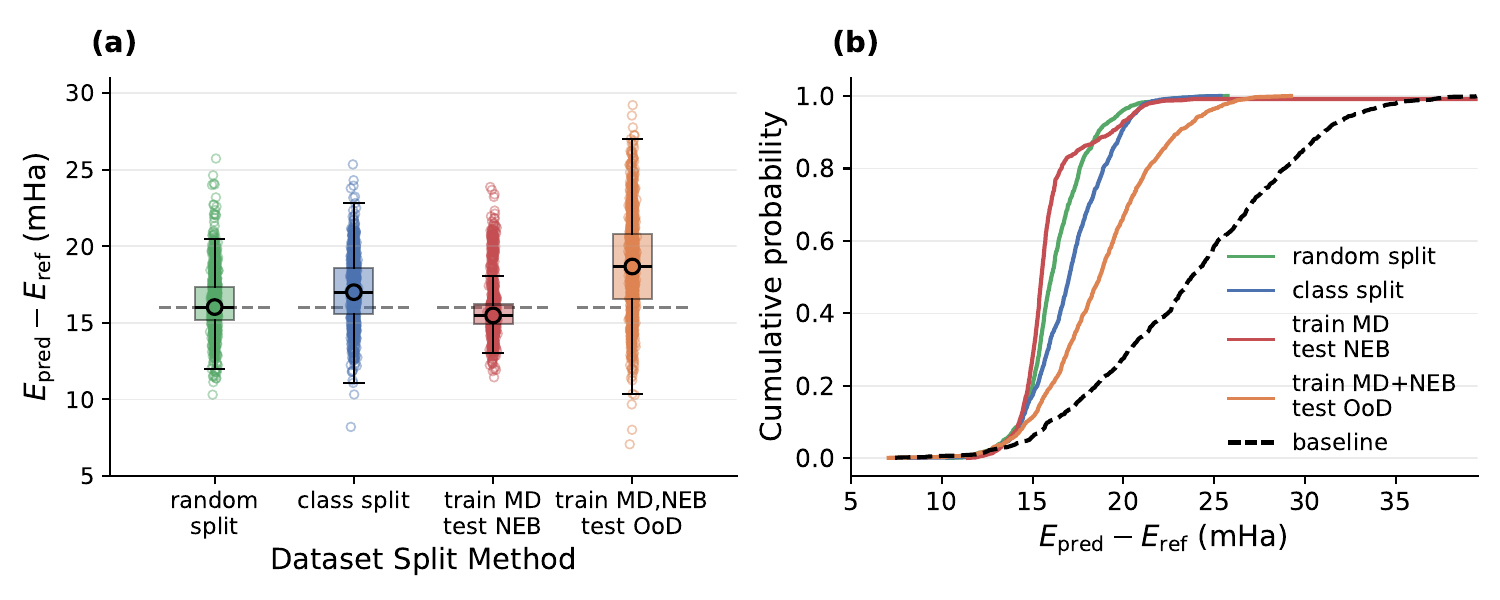}
    \caption{\textbf{Generalization performance across conformer classes and out-of-distribution conformational regimes.} \textbf{(a)} Energy error for the 14-qubit $\varepsilon=16$~mHa dataset across different methods for dividing the train and test datasets. \textbf{(b)}  Cumulative probability distributions for the energy error with respect to the CASCI energy.
    }
    \label{fig:14q-molecular-generalization}
\end{figure}

For the main-text results, we randomly split the dataset into training, validation, and test sets, so that the reported metrics reflect performance on samples drawn from the same distribution as the training data. However, this procedure can place highly similar structures in both the training and test sets. Since molecular dynamics trajectories contain substantial redundancy from correlations between nearby snapshots and repeated sampling of the same regions of conformational space, a random split can introduce some degree of information leakage into the test set. We provide a set of more stringent tests of molecular generalization in Figure~\ref{fig:14q-molecular-generalization}. Results in this section are provided for the Gemma model with pretraining only on the 14-qubit $\varepsilon=16$~mHa dataset. To standardize the results in this section across dataset subsets of different sizes, for all comparisons shown here we sample $12,000$ structures for the training set and $950$ for the test set.

\begin{itemize}
\item \textbf{Generalization to unseen conformer classes --} As described in section~\ref{subsec:DataGen} of the main text, we calculate 15 reference conformers of imipramine using the ETKDG algorithm. These reference conformers are chemically-plausible geometries that represent distinct regions of conformational space. We label each structure in the dataset by the closest of the 15 reference conformers, where closeness is defined by the minimum dihedral-angle difference. We then use the structures labelled by the 3 highest-energy reference conformers as the test set, and use rest of the structures for the training and validation sets. However, we find that dividing the structures into class labels by the similarity to these reference conformers is highly ambiguous, with several reference conformers often having very similar dihedral angle differences from the same structure. This suggests that this split may reduce, but not entirely eliminate, the possibility of leakage between the training and test sets.

\item \textbf{Generalization between molecular dynamics and transition states --} The reference conformers described in the previous section are not themselves members of the dataset, although they are intended to be representative. Similarly, as described in Section~\ref{subsec:DataGen}, the NEB transition states are sampled from transition pathways between reference conformers and provide representative, out-of-dataset structures. We therefore train the model on molecular dynamics snapshots and reserve the NEB transition states for testing.

\item \textbf{Generalization to out-of-distribution configurations --} As described in Section~\ref{subsec:DataGen}, we generate a dataset of high-energy structures with energies that make them not accessible by MD simulations and NEB pathways. 
We train the model on molecular dynamics simulations and NEB transition states and retain only out-of-distribution conformers for the test set. We consider this our most stringent test of molecular generalization, since we can guarantee that no similar structures are shared between the train and test set, and the test set structures come from a higher-energy distribution that any found in the training set.

\end{itemize}
 
Fig. \ref{fig:14q-molecular-generalization} shows the generalization capabilities of the Gemma model by evaluating the energy error across these three distinct out-of-distribution regimes. We find that model accuracy is slightly higher when training on MD simulations and testing on NEB than for a random split over the full dataset. This suggests that the NEB transition states are comparatively easy for the model to predict relative to other structures in the dataset. Energy accuracy is marginally lower for the split by conformer class label than for the random split. This effect is amplified for testing on out-of-distribution conformers, where the generated circuit accuracy degrades noticeably relative to the random split.

To facilitate interpreting these results, we provide a baseline for the expected circuit generation accuracy if the model reproduced elements of the training dataset perfectly but did not learn how to use information about an individual molecule to condition circuit generation. Specifically, we show the energy accuracy arising from evaluating sequences in the test set on Hamiltonians from random (different) structures from the training set. The substantially better model accuracy on the out-of-distribution conformer test set relative to this baseline implies that the model is learning generalizable things about circuit generation that it can apply to out-of-distribution molecular systems, albeit with somewhat reduced precision.

\subsubsection{Non-Hamiltonian molecular representations}
\label{appendix_sub:molecular_representation}

\begin{figure}[htbp]
    \centering
    \includegraphics[width=\textwidth]{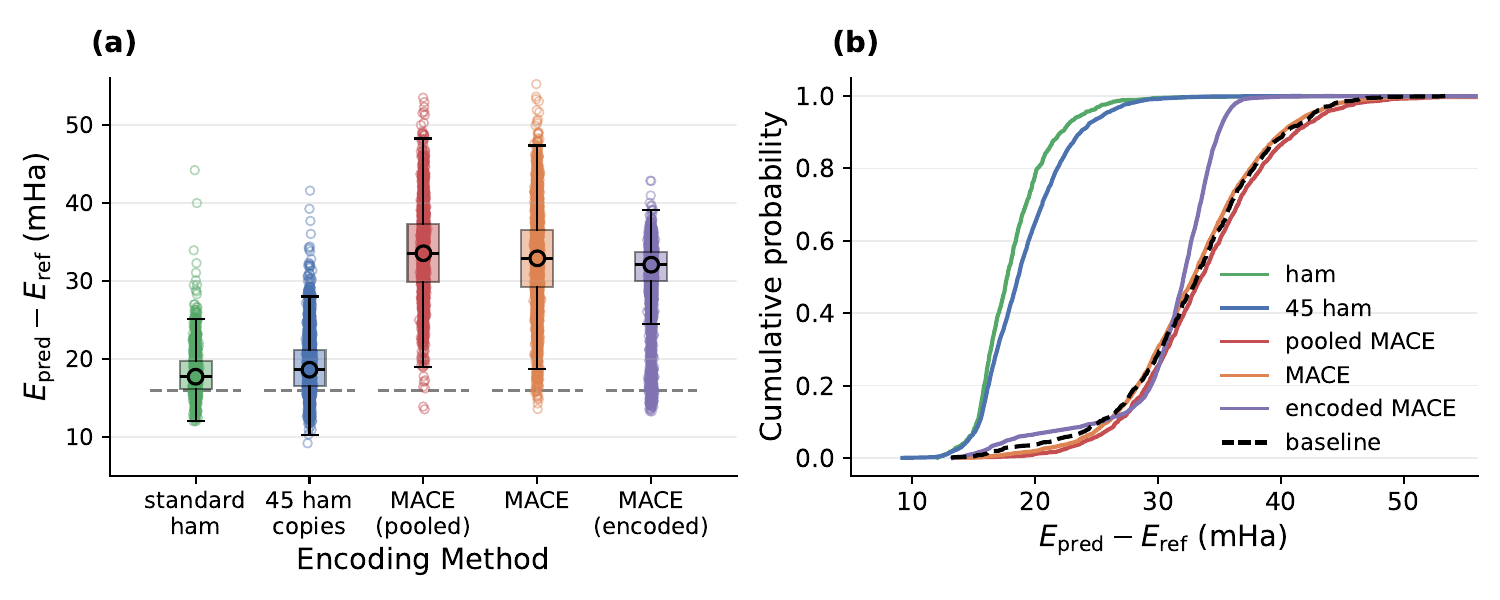}
    \caption{\textbf{Model performance across different methods for molecular encoding.} \textbf{(a)} Energy error for the 14-qubit $\varepsilon=16$~mHa dataset across different methods for the molecular encoding (results are for one-off sequence generation rather than best-of-16). \textbf{(b)} Cumulative probability distributions for the energy error with respect to the reference. We find substantially better model performance using Hamiltonian coefficients to input molecular information than providing geometrical information with MACE embeddings.}
    \label{fig:MACE}
\end{figure}

The models described in this work accept information about a specific molecular configuration through a special multimodal token, and use that information to generate high-quality ground state circuits for that particular configuration. In the main text we describe a strategy to canonically-order Pauli strings in the molecular Hamiltonian, and provide the model with a vector of coefficients of each of those strings. However this has the disadvantage that the coefficient vectors do not have any generalizability to different active spaces or different molecules. 

To address this deficiency, we have explored providing conformer information via the learned feature embedding extracted from MACE-OFF~\cite{mace-off}. MACE-OFF is a pretrained MACE-based foundation model for organic molecular force fields. Before output feature extraction, MACE-OFF embeds the 3-dimensional coordinates of atoms in a molecule into a learned feature embedding of dimension $(n_\text{atoms}, n_\text{features})$. Imipramine has $45$ atoms and we use the ``medium'' MACE-OFF model, for which $n_\text{features}=640$. In this section we assess the utility of the MACE-OFF embedding for conditioning ground state circuit generation.

Since the MACE-OFF feature embeddings are much larger than the Hamiltonian coefficient vectors, we consider several ways to process them for input to the model.
\begin{itemize}
    \item \textbf{Pooling -- } the full MACE-OFF feature embedding is pooled along the atom dimension, giving rise to a vector of length $n_\text{features}$ that is fed into the encoder and inserted at the special multimodal token (in the same way as the Hamiltonian coefficients).
    \item \textbf{Encoded embeddings -- } MACE-OFF feature embeddings are provided at the input to the Hamiltonian encoder. The encoder acts simultaneously to encode the feature vector of each atom. Results are then pooled after the encoder and inserted at the special multimodal token.
    \item \textbf{Full MACE-OFF embeddings -- } the full MACE-OFF feature embeddings are encoded and then are provided directly to the model as $n_\text{atoms}$ separate multimodal tokens.
\end{itemize}
To aid interpretation, we introduce a baseline that estimates the circuit-generation accuracy expected if the model simply reproduced training-set circuits, without learning to condition its outputs on the target molecule. Specifically, we evaluate test-set operator sequences on Hamiltonians from randomly selected, distinct structures in the training set and report the resulting energy accuracy. 
We additionally perform an ablation study in which the model is given $45$ repeated copies of the Hamiltonian coefficients. This controls for sequence length and tests whether the difficulty of using the MACE-OFF embedding for circuit generation is primarily due to its larger number of input tokens.

Figure~\ref{fig:MACE} compares the energy errors obtained with different molecular encoding strategies after pretraining of the Gemma model. Overall, the MACE-based embeddings lead to substantially worse performance than the Hamiltonian coefficients. As shown in Figure~\ref{fig:MACE}b, the full and pooled MACE-OFF embeddings perform comparably to the baseline, indicating that the model does not effectively use these representations to condition circuit generation on the target molecule. The encoded MACE-OFF representation yields slightly improved results, suggesting that the model can extract some molecule-specific information from these features. However, for the same training time, this approach remains much less effective than providing the Hamiltonian coefficients directly. As a control, we provide the model with 45 copies of the Hamiltonian coefficients, matching the larger token count of the MACE-OFF embeddings. The resulting performance, which is somewhat worse but comparable to the training with a single vector of Hamiltonian coefficients, confirms that the degradation in accuracy using the MACE-OFF is not caused by sequence length alone.

\subsubsection{Model size dependence for the Gemma model}

\begin{figure}[htbp]
    \centering
    \includegraphics[width=0.9\textwidth]{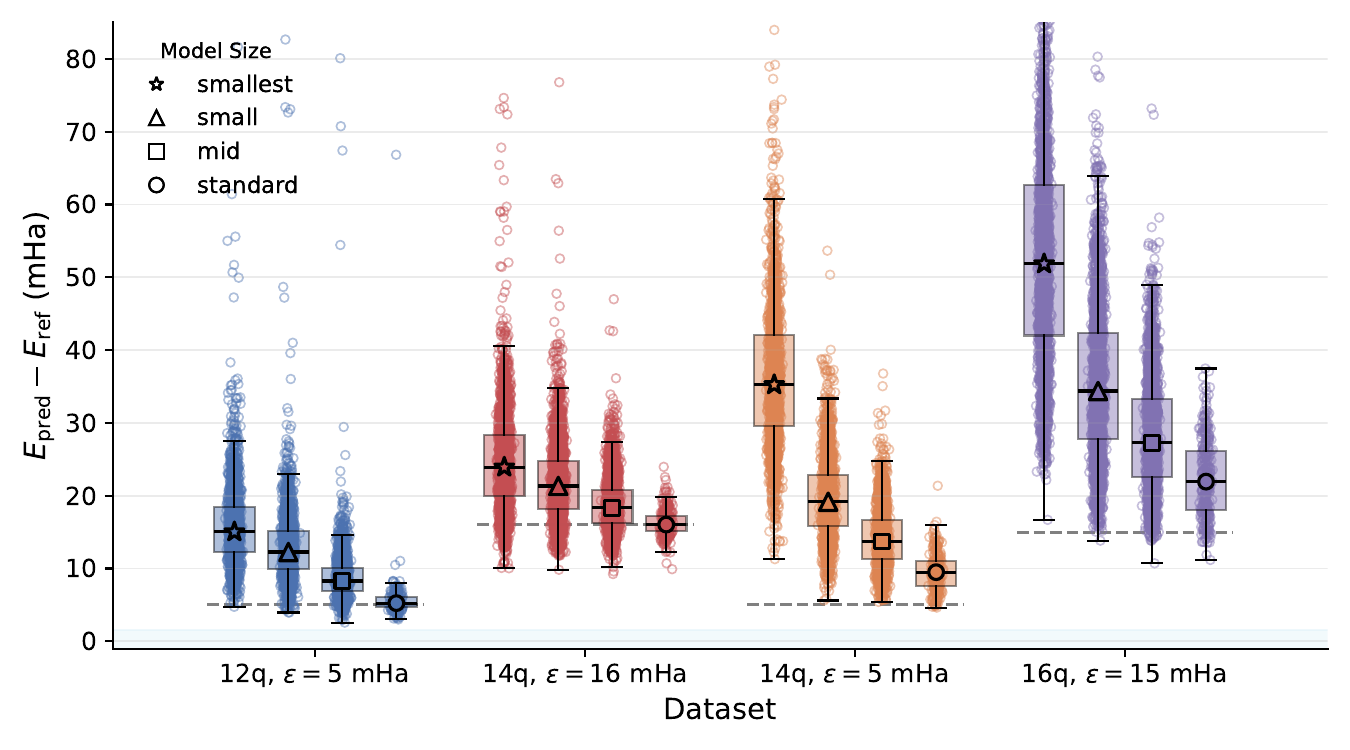}
    \caption{\textbf{Circuit generation accuracy for increasing model size.} We show the performance of the Gemma model after pretraining only for different model sizes. Circuit generation quality saturated at the training dataset accuracy for the two simpler datasets, 12-qubit $\varepsilon=5$~mHa and 14-qubit $\varepsilon=16$~mHa. For the two more complex datasets, results suggest that some moderate gains in pretraining accuracy could still be achieved by further increasing the model size.} 
    \label{fig:model_size_dep}
\end{figure}

Given that the Gemma model is trained from scratch on a relatively small, task-specific dataset, we aim to use the smallest model size that still achieves strong performance on our task. 
Smaller models have the benefit of faster inference time and more efficient training, which is especially important for RL. To assess how circuit-generation accuracy depends on model size, we perform an ablation study across several model sizes and evaluate performance after pretraining alone. 

As discussed in the main text, the 12-qubit $\varepsilon=5$~mHa and 14-qubit $\varepsilon=16$~mHa datasets are simpler in terms of model performance because they typically have relatively short sequence lengths. For these datasets, we consider model sizes of $11$M, $22$M, $100$M, and $325$M (``standard''). For the more complex 14-qubit $\varepsilon=5$~mHa and 16-qubit $\varepsilon=15$~mHa datasets, we consider models with reduced sizes of $22$M, $100$M, and $325$M. The ``standard'' model sizes from the main text are $474$M for the 14-qubit $\varepsilon=5$~mHa dataset and $508$M for the 16-qubit dataset. For the simple datasets we find that the model sizes used in the main text are sufficient to saturate the pretraining accuracy at the accuracy threshold of the ADAPT-VQE training data. For the more complex datasets, we see some evidence that even larger models could give a moderate improvement in pretraining accuracy.

\subsubsection{Alternate reward function for fine-tuning}
\label{appendix_sub:reward_function}

\begin{figure}[htbp]
    \centering
    \includegraphics[width=0.8\textwidth]{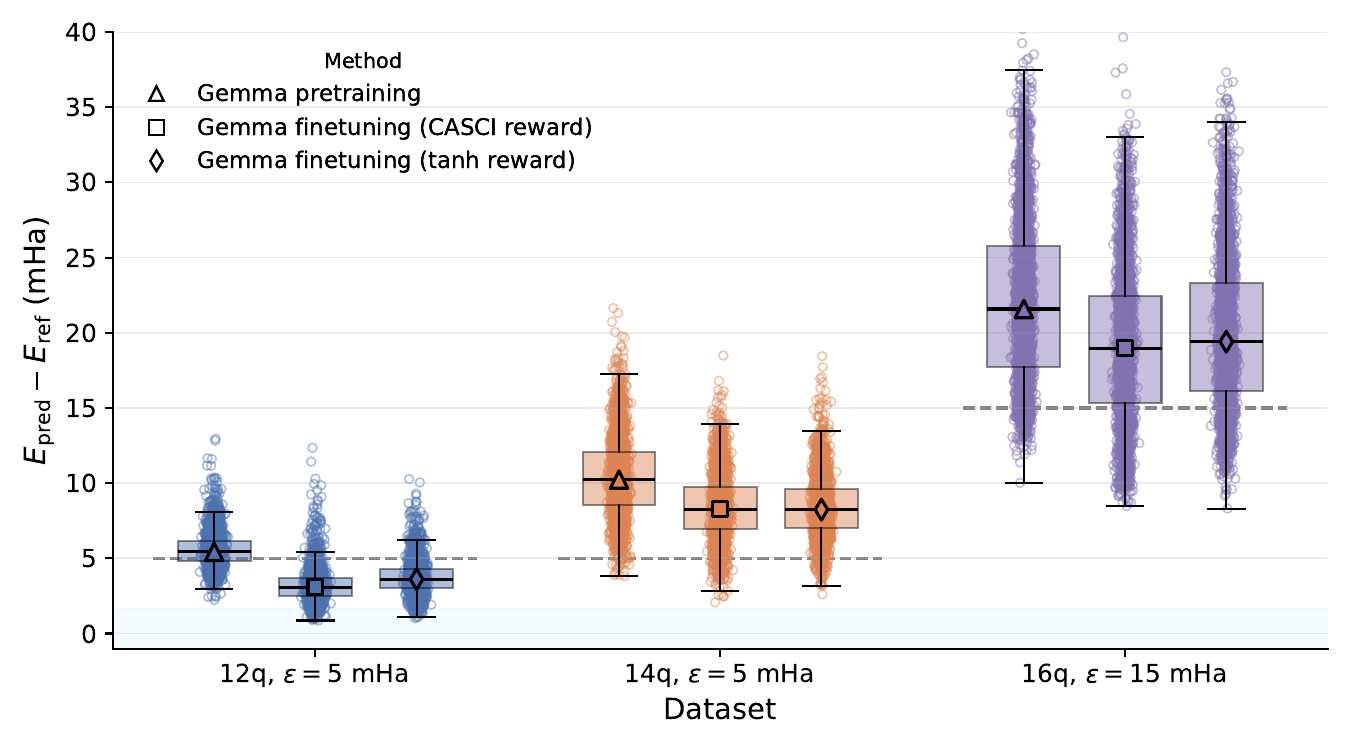}
    \caption{\textbf{Circuit generation accuracy after fine-tuning with different reward functions.}
    Energy accuracy of generated circuits from the Gemma model for the most accurate datasets at each qubit number after pretraining (triangles) and after fine-tuning with the main text reward function (Equation~\ref{eq:casci_reward}, squares) and the $\tanh$ reward function described here (diamonds). 
    }
    \label{fig:reward_functions}
\end{figure}

The primary reward function used for fine-tuning the Gemma model in the main text uses a reference energy (CASCI or CCSD) as a target, and strongly increases the reward for model-generated circuits that achieve energies approaching this value. We find this formulation to be effective when a suitable reference energy is available, but computing the CASCI energy can be computationally expensive for large active spaces. In this section we demonstrate another reward function which can achieve comparable performance without requiring a reference energy.

We choose a reward function of the form
    \begin{equation}
        R = \tanh(\alpha(E_{\text{HF}} - E_\text{circuit})) \label{eq: tanh reward}
    \end{equation}
where $\alpha$ is a tunable parameter that controls the steepness of the reward. Smaller $\alpha$ provides a gentler reward that encourages exploration, while larger $\alpha$ encourages exploitation. The function $\tanh(\Delta E)$ is approximately linear for small $\Delta E$ while mapping energy differences to the range $(-1,1)$, avoiding the hard clipping in Equation~\ref{eq:casci_reward} and providing a reward signal less sensitive to large energy differences within a single generation group. In practice, $\alpha$ is calculated dynamically during subsequent fine-tuning iterations to ensure that most generated circuits remain in the approximately linear regime of the $\tanh$.

Figure~\ref{fig:reward_functions} shows the energy accuracy for the Gemma model fine-tuned with this reward function compared to fine-tuned with Equation~\ref{eq:casci_reward}. We find the reward function in Equation~\ref{eq:casci_reward} to be moderately more effective in terms of accuracy for the same training time, with the reward function using the CASCI energy having an improvement of $0.5$~mHa for the 12-qubit dataset, no improvement for the 14-qubit dataset, and an improvement of $0.7$~mHa for the 16-qubit dataset. These results suggest that incorporating a reference energy into the reward function provides some benefit, but is not critical. This is encouraging for future applications in which obtaining such a reference energy may be computationally expensive.

\subsubsection{Operator Counts and Resource Estimation}
\label{appendix_sub:ciruit_resource_requirements}

\begin{figure}[htbp]
    \centering
    \includegraphics[width=\textwidth]{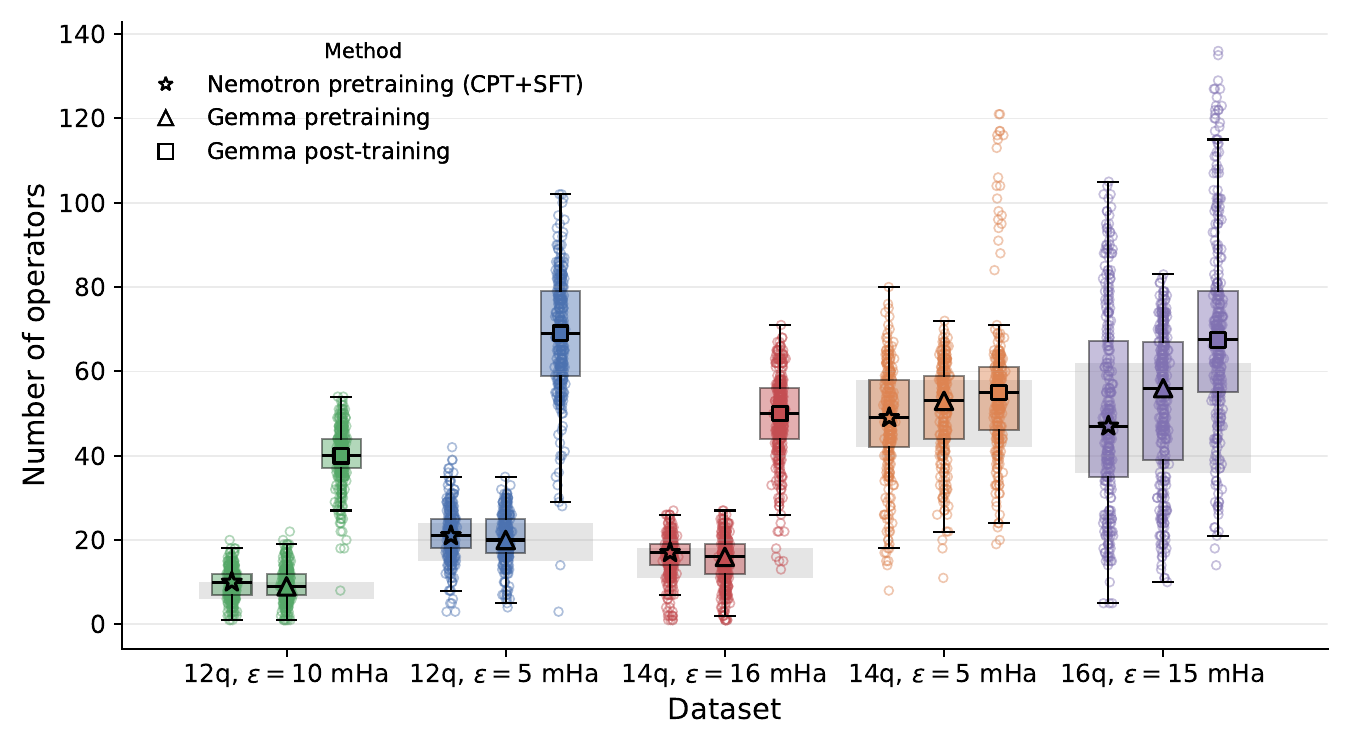}
    \caption{\textbf{Operator counts in generated sequences.} Number of operators selected from the pool by Nemotron and Gemma models after pretraining (stars and triangles) and Gemma after RL fine-tuning (squares). The interquartile range of the number of operators in the ADAPT-VQE training datasets are shown as shaded grey boxes for reference. Generally models achieve better accuracy by increasing the number of selected operators.}
    \label{fig:operator-count}
\end{figure}

In Figure \ref{fig:operator-count}, we show the number of operators in circuits generated by the Nemotron model after pretraining, and the Gemma model both after pretraining only and after post-training fine-tuning. The interquartile range of the operator counts from ADAPT-VQE datasets are shown as shaded grey boxes. For both Nemotron and Gemma models after pretraining only, model-generated circuits have very similar numbers of operators to the ADAPT-VQE training datasets. Moderate increases in the operator counts may be due to the best-of-16 selection criteria, which moderately favors longer sequences if they can produce better energies. Post-training fine-tuning dramatically increases the number of operators for both 12-qubit datasets and the higher-tolerance 14-qubit dataset. These yield generated circuits with much better accuracy than the training datasets, as shown in Figure~\ref{fig:mean-energy}. For the two most complex datasets, 14-qubit $\varepsilon=5$~mHa and 16-qubit $\varepsilon=15$~mHa, we find that fine-tuning does not significantly increase the number of operators beyond the training set, and the circuit generation accuracy also does not improve significantly beyond the accuracy of the ADAPT-VQE dataset. We speculate that it may be more challenging for RL to effectively propose longer sequences for these datasets and may therefore require longer training.

\begin{figure}
    \centering
    \subfloat[Canonical two-qubit gate counts]{\includegraphics[width=0.7\textwidth]{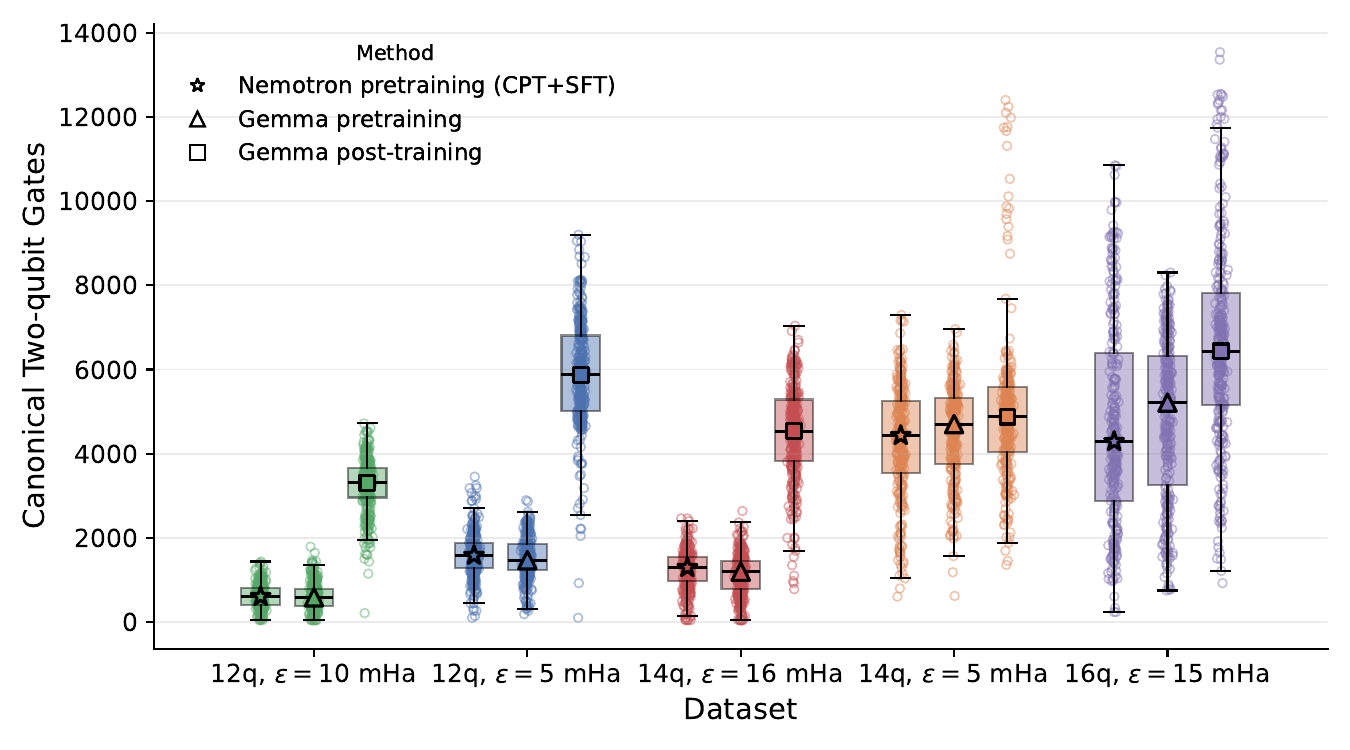}}\\
    \subfloat[Optimized two-qubit gate counts]{\includegraphics[width=0.7\textwidth]{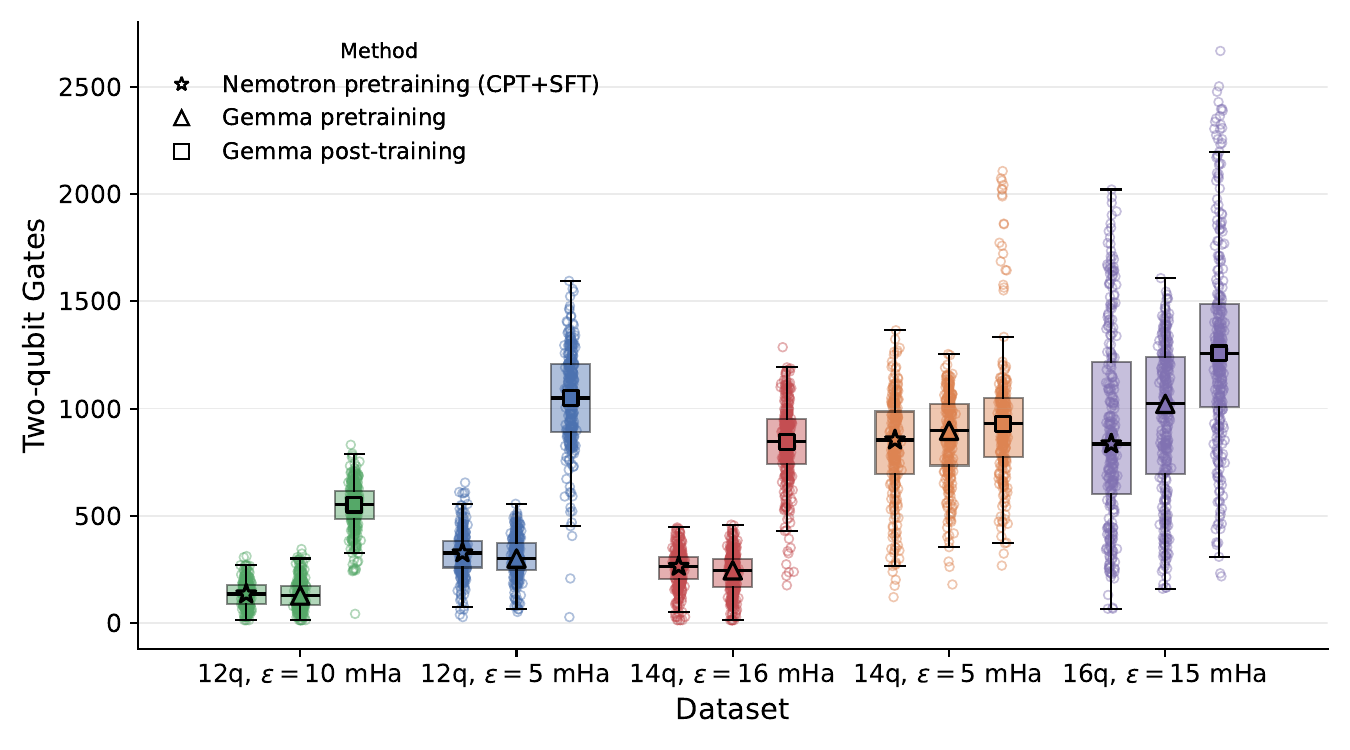}}\,
    \caption{\textbf{Canonical and optimized two-qubit gate counts for model generated circuits.} 
    \textbf{(a)} Two-qubit gate counts in the canonical representation (using CNOT and rotation gates). \textbf{(b)} Two-qubit gate counts for optimized circuits compiled with \texttt{pytket} for Quantinuum Helios-1.
    }
    \label{fig:gemma_nemo_circuit_resources}
\end{figure}

To supplement the results on the fermionic operator count for model-generated circuits, we additionally provide both canonical and optimized two-qubit gate counts for these sequences in Figure \ref{fig:gemma_nemo_circuit_resources}.

\subsection{Molecular Data Generation}
\label{appendix:molecular_data_generation}

In this section we provide supplemental details on the dataset generation, characteristics, and dataset diversity.

\subsubsection{Dataset Generation}
\label{appendix_sub:dataset}

The initial geometry for imipramine used to initialize MD simulations was pre-optimized with the Lennard-Jones potential and subsequently optimized using the pretrained “small” variant of the MACE-OFF machine learning potential \cite{Batatia2022Design, Batatia2022mace, mace-off}. The full geometry is given in  Figure~\ref{lst:imipramine}.
All simulations were performed in gas phase. Our dataset is composed of 5 independent MD simulations with different seeds for choosing the initial velocity distribution. For 4 of those simulations, one frame was selected every ten saved frames. For the fifth we used a mixed sampling strategy and selected frames at intervals of five and ten saved frames.

We identified 15 reference conformers of imipramine using the random distance matrix method implemented in RDKIT. NEB calculations provided in the main text are carried out between every pair among these reference conformers. Each NEB calculation constructs a transition pathway connecting a pair of structures. Pathways are computed using the Atomic Simulation Environment (ASE) package \cite{ase-paper} and each NEB path is discretised into 30 intermediate images. Energies and forces along the pathways are evaluated using the MACE-OFF potential. For 17 of the 105 conformer pairs, NEB pathways fail to converge in a fixed time window or exhibit unphysical bond distortions or excessively high energies. These are discarded and images from the remaining 88 NEB pathways are incorporated in the dataset.

We additionally identified a set of 122 reference conformers using CREST~\cite{CREST}, a multi-step workflow including meta-dynamics simulations for exploring conformer space. The highest-energy of these was perturbed with the rattle method, as described in the main text, to form a set of configurations that are out-of-distribution with respect to MD simulations and NEB transition states.

\begin{figure}[t]
\centering
\begin{multicols}{2}
\begin{lstlisting}[style=xyzstyle]
45
N      -2.67643818       2.44912632       2.58767414
N       1.58709562       1.83667298       2.87736438
C      -1.28549061       2.09858610       2.23956342
C      -3.74399016       1.50359424       2.34929173
C      -3.08369177       3.77939590       2.23254831
C      -5.24784529       2.63298578       4.19235608
C      -4.24051318       3.77415863       4.43927654
C      -4.93537096       1.56898319       3.13542631
C      -3.86084480       4.47673681       3.16208513
C      -0.76115045       0.99657998       3.17063898
C       0.71129406       0.64299041       2.89500142
C      -3.65754465       0.53842571       1.32033822
C      -2.78515483       4.36040468       0.98745540
C      -5.95222489       0.62417094       2.90380927
C      -4.29084373       5.78115097       2.87638491
C      -4.69087986      -0.37265490       1.10616230
C      -3.21982990       5.65910014       0.70470854
C      -5.82986246      -0.33781589       1.90314659
C      -3.96442550       6.37127272       1.65083354
C       2.89391532       1.48941323       2.29669455
C       1.78163621       2.38246737       4.23243248
H      -1.18333730       1.80631271       1.17464454
H      -0.64619114       2.99739202       2.38337117
H      -6.21605774       3.10782795       3.91649544
H      -5.41164878       2.11466330       5.16269950
H      -3.32963917       3.37325620       4.93546001
H      -4.69763997       4.49595299       5.15173166
H      -1.37053645       0.07716900       3.03384537
H      -0.88647488       1.31169535       4.22853085
H       0.74434753       0.13265939       1.90580562
H       1.06005419      -0.09353487       3.65389703
H      -2.80124632       0.47752812       0.66701224
H      -2.21949161       3.81487086       0.24367404
H      -6.86350549       0.64900146       3.49083779
H      -4.88756000       6.33236078       3.59213199
H      -4.60856230      -1.10813208       0.31599989
H      -2.98246861       6.11259339      -0.24901364
H      -6.63043312      -1.04669620       1.73411705
H      -4.30037737       7.37611204       1.42913211
H       2.76792879       1.12777397       1.25343571
H       3.40401008       0.70207207       2.89583468
H       3.54531604       2.38890128       2.25022561
H       2.48087392       3.24632707       4.20273283
H       2.18829263       1.61402713       4.92663375
H       0.82423818       2.77277967       4.63437779
\end{lstlisting}
\end{multicols}
\caption{\textbf{Imipramine geometry used to initialize MD simulations.} Atom types and 3-dimensional coordinates of the imipramine molecule used to initialize MD simulations.}
\label{lst:imipramine}
\end{figure}

In Table~\ref{tab:datasets} we summarize the size of each dataset generated for the results in the main text, including the percentages of the dataset generated from molecular dynamics simulations, sampled from NEB transition pathways, and from the out-of-distribution configurations.

\begin{table}[ht]
\centering
\begin{tabular}{cccccccc}
\toprule
Qubits & Tolerance (mHa) & Operator Pool & Total Size & MD (\%) & NEB (\%) & OoD(\%) \\
\midrule
12 & 10 & UCCGSD & 15,475 & 76.3 & 17.2 & 6.5 \\
12 & 5 & UCCGSD & 14,332  & 77.4 & 15.6 &  7.0 \\
14 & 16 & UCCGSD & 15,697 & 77.0 & 16.6 & 6.4 \\
14 & 5 & UCCGSD & 12,997 & 89.2 & 7.2 & 3.6 \\
16 & 15 & UCCSD & 15,498 & 85.5 & 8.1 & 6.4 \\
\bottomrule
\end{tabular}
\caption{\textbf{Characteristics of the datasets considered in this work.}}
\label{tab:datasets}
\end{table}

\subsection{Dataset Diversity}
\label{appendix_sub:data_diversity}

\begin{figure}[htbp]
    \centering
    \includegraphics[width=\textwidth]{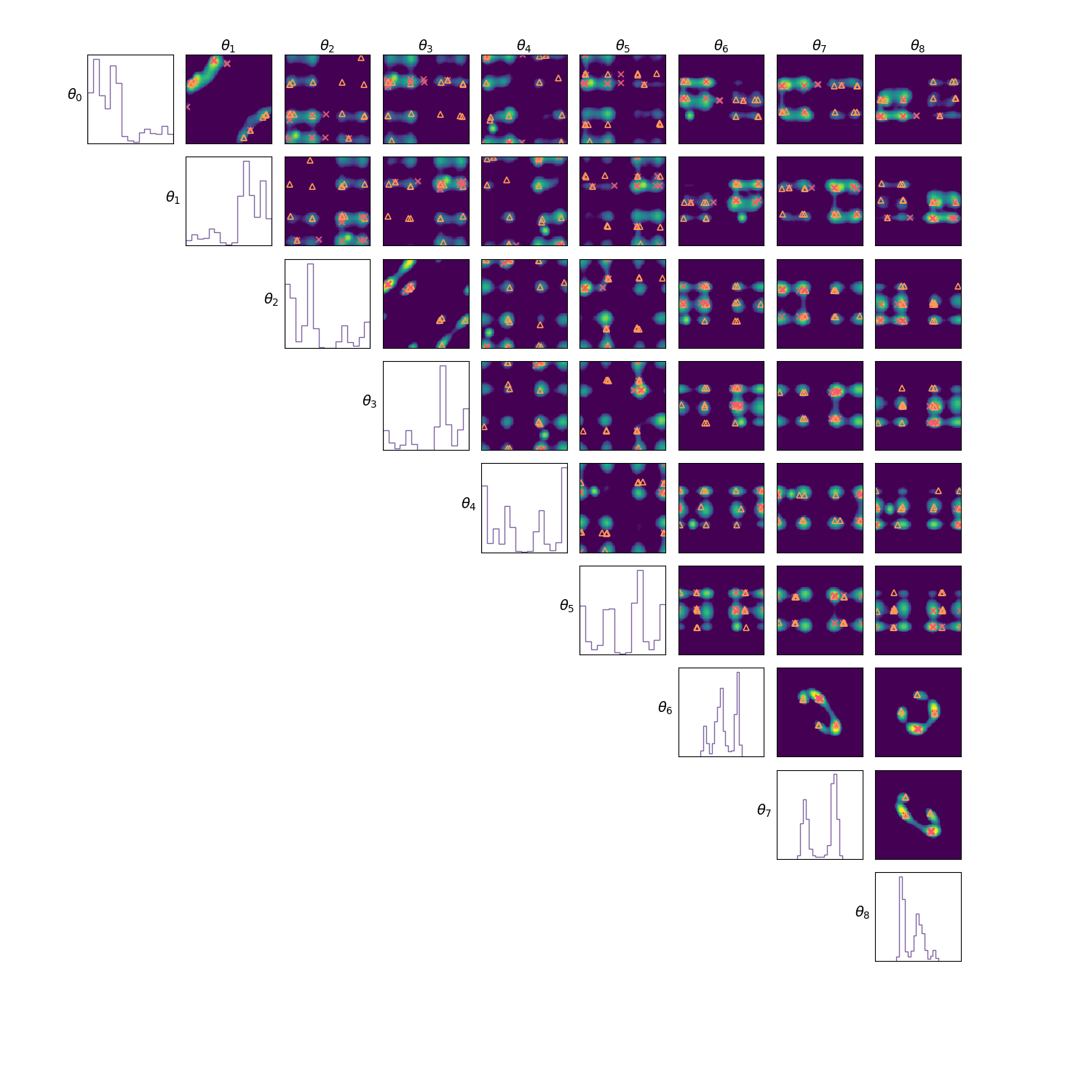}
    \caption{\textbf{Angular correlations for all dihedral angles.} All axes range from $[-\pi, \pi]$ and lighter colors indicate higher dataset density. Labels are excluded for visual clarity.}
    \label{fig:full_dihedral}
\end{figure}

In the main text, we illustrated the diversity of the generated dataset by showing the correlation of a representation dihedral angle $\theta_0$ with all other dihedral angles. In Figure~\ref{fig:full_dihedral} we supplement those results with the full correlations between all dihedral angles. These reinforce that the dataset spans a diverse set of structures covering allowed ranges in the dihedral angle space, and reflect comparable diversity to the reference conformer sets obtained by independent methods.

\subsection{Molecular Encoder}
\label{appendix:encoder}

We represent the Hamiltonian as a dense vector $\mathbf{h} \in \mathbb{R}^{d}$ normalized via MaxAbs scaling $\hat{h} = \frac{h_i}{\max_j |h_j|}$ which maps all coefficients to $[-1, 1]$ while preserving their relative structure. A residual MLP encoder $f_\text{enc}: \mathbb{R}^{d} \to \mathbb{R}^{D}$ then maps the normalized Hamiltonian vector to the transformer hidden dimension $D$. The encoder is structured as follows: an input LayerNorm over the $d$-dimensional input; a linear projection from $d$ to an internal hidden dimension $H$; $L$ residual MLP blocks, each applying pre-norm LayerNorm, a two-layer feedforward network with SiLU activation and expansion factor $\alpha$
(mapping the hidden dimensions $H \to \alpha H \to H$),  dropout, and an additive residual connection; a final LayerNorm over $H$; and a linear output projection from $H$ to $D$. Only the input LayerNorm and input projection depend on the problem-specific dimension $d$ and all residual blocks and output layers are structurally-identical across problem sizes.

The encoder hyperparameters are chosen per problem size to balance representational capacity with computational cost. Table \ref{tab:encoder_params} summarizes the encoder configuration for each qubit count studied in this work. 

\begin{table}[ht]
\centering
\begin{tabular}{lccc}
\toprule
Parameter & \textbf{12 qubits} & \textbf{14 qubits} & \textbf{16 qubits}\\\hline
Hamiltonian dim $d$ & 1819 & 3382 & 5793 \\
Hidden dim $H$ & 2048 & 3072 & 4096 \\
Depth $L$ & 4 & 4& 4 \\
FFN multplier $\alpha$ & 2.0 & 1.5 & 1.0\\
Dropout & 0.2 & 0.1 ($\varepsilon=5$~mHa); 0.2 ($\varepsilon=16$~mHa) &0.2 \\
\bottomrule
\end{tabular}
\caption{Hamiltonian encoder configuration for each problem size. The FFN multiplier $\alpha$ is reduced at larger qubit counts to keep the intermediate expansion $(\alpha H)$ tractable given the increased input dimension. Encoder parameters are the same for both models, except that the output dimension $D$ matches the respective backbone: 5120 for Nemotron and 1024 for Gemma.}
\label{tab:encoder_params}
\end{table}

\subsection{Nemotron architecture details}
\label{appendix:nemotron_architecture}

\subsubsection{Chat template}
\label{appendix_sub:nemotron_tokenization}
To insert molecular information into the Nemotron model, the special Hamiltonian insertion token is placed at a fixed position within an instruction-following prompt template:
\begin{verbatim}
System: You generate quantum circuits as (index, theta) pairs.
User:   You are given a Hamiltonian. Generate the ADAPT-VQE circuit
as one `<index> <theta>' pair per line.
Hamiltonian: <ham>
\end{verbatim}
During the forward pass, the encoder output replaces the standard token embedding at the \texttt{<ham>} position in-place before entering the transformer, so that it is processed by all subsequent Mamba2, attention, and MLP layers.

\subsubsection{Hyperparameters}
\label{appendix_sub:nemotron_parameters}

All Nemotron training experiments use a cosine learning rate schedule with warmup, the AdamW optimizer~\cite{loshchilov2019decoupledweightdecayregularization}, weight decay, gradient checkpointing, BF16 mixed precision, and DeepSpeed ZeRO Stage 3~\cite{rajbhandari2020zeromemoryoptimizationstraining} for distributed multi-GPU training. Best checkpoints are selected by validation loss and SFT is initialized from the best CPT checkpoint. Response-only masking is applied (train on completion tokens only). Hyperparameters are detailed in Tables~\ref{tab:cpt_hyperparams_Nemotron} and \ref{tab:sft_hyperparams_Nemotron}.

\begin{table*}[ht]
\centering
\caption{\textbf{Continual pretraining (CPT) hyperparameters for each dataset.}}
\label{tab:cpt_hyperparams_Nemotron}
\resizebox{\textwidth}{!}{%
\begin{tabular}{lcc|cc|c}
\toprule
& \multicolumn{2}{c}{\textbf{12 qubits}} & \multicolumn{2}{c}{\textbf{14 qubits}} & \textbf{16 qubits}
\\
\textbf{Hyperparameter} & $\varepsilon=10$~mHa & $\varepsilon=5$~mHa & $\varepsilon=16$~mHa & $\varepsilon=5$~mHa & $\varepsilon=15$~mHa \\
\midrule
Learning rate           & $2\times10^{-5}$ & $2\times10^{-5}$ & $2\times10^{-5}$ & $2\times10^{-5}$ & $2\times10^{-5}$ \\
Weight decay            & 0.01             & 0.01             & 0.05             & 0.01             & 0.01             \\
Warmup ratio            & 0.05             & 0.05             & 0.10             & 0.05             & 0.05             \\
Epochs                  & 3                & 5                & 4                & 4                & 5                \\
Per-device batch size   & 1                & 1                & 1                & 1                & 1                \\
Gradient accumulation   & 1                & 1                & 1                & 1                & 1                \\
Effective batch size    & 20               & 20               & 20               & 20               & 20               \\
Number of GPUs          & 20               & 20               & 20               & 20               & 20               \\
Max sequence length     & 1024             & 1024             & 1024             & 1024             & 1280             \\
Max gradient norm       & 1.0              & 1.0              & 1.0              & 1.0              & 1.0              \\
\bottomrule
\end{tabular}
}
\end{table*}

\begin{table*}[ht]
\centering
\caption{\textbf{Supervised fine-tuning (SFT) hyperparameters for each dataset.}}
\label{tab:sft_hyperparams_Nemotron}
\resizebox{\textwidth}{!}{%
\begin{tabular}{lcc|cc|c}
\toprule
& \multicolumn{2}{c}{\textbf{12 qubits}} & \multicolumn{2}{c}{\textbf{14 qubits}} & \textbf{16 qubits}
\\
\textbf{Hyperparameter} & $\varepsilon=10$~mHa & $\varepsilon=5$~mHa & $\varepsilon=16$~mHa & $\varepsilon=5$~mHa & $\varepsilon=15$~mHa \\
\midrule
Learning rate           & $3\times10^{-6}$ & $3\times10^{-6}$ & $5\times10^{-6}$ & $3\times10^{-6}$ & $3\times10^{-6}$ \\
Weight decay            & 0.05             & 0.05             & 0.05             & 0.05             & 0.05             \\
Warmup ratio            & 0.10             & 0.10             & 0.10             & 0.10             & 0.10             \\
Epochs                  & 6                & 6                & 4                & 6                & 6                \\
Per-device batch size   & 1                & 1                & 1                & 1                & 1                \\
Gradient accumulation   & 2                & 2                & 2               & 2                & 2                \\
Effective batch size    & 40               & 40               & 40               & 40               & 40               \\
Number of GPUs          & 20               & 20               & 20               & 20               & 20               \\
Max sequence length     & 1024             & 1024             & 1024             & 1024             & 1280             \\
Max gradient norm       & 1.0              & 1.0              & 1.0              & 1.0              & 1.0              \\
Group by length & False & False & True & False & False \\
Early stopping patience & 6 & 6 & 3 & 3 & 6 \\
\bottomrule
\end{tabular}
}
\end{table*}

\subsection{Gemma architecture details}
\label{appendix:gemma_architecture}

\subsubsection{Hyperparameters}
\label{appendix_sub:gemma_parameters}

All Gemma~3 training experiments use a cosine learning rate schedule with warmup, AdamW~\cite{loshchilov2019decoupledweightdecayregularization}, weight decay and dropout. During RL, we accelerate model generations using vLLM~\cite{kwon2023efficientmemorymanagementlarge} colocated on the same GPU as the GRPO trainer process. The hyperparameters for the Gemma model during pretraining and post-training fine-tuning for each dataset are shown in Table~\ref{tab:pretrain_hparams} and Table~\ref{tab:rl_hparams}, respectively.

\begin{table}[h]
  \caption{\textbf{Pretraining hyperparameters for each qubit count and tolerance.}}
  \label{tab:pretrain_hparams}
  \centering
  \begin{small}
  \setlength{\tabcolsep}{4pt}
  \begin{tabular}{lccccc}
    \toprule

      & \multicolumn{2}{c}{\textbf{12 qubits}}
      & \multicolumn{2}{c}{\textbf{14 qubits}}
      & \textbf{16 qubits} \\
    \cmidrule(lr){2-3}
    \cmidrule(lr){4-5}
    \cmidrule(lr){6-6}
    \textbf{Tolerance}
      & $0.005$ & $0.01$
      & $0.005$ & $0.016$
      & $0.015$ \\
    \midrule
    \multicolumn{6}{l}{\textbf{Gemma 3}} \\
    Decoder hidden size
      & 1024 & 1024 & 1024 & 1024 & 1024 \\
    Number of layers
      & 16 & 16 & 20 & 16 & 20 \\
    Attention heads
      & 8 & 8 & 12 & 8 & 12 \\
    Context length
      & 1024 & 1024 & 2048 & 1024 & 2048 \\
    Sliding window
      & 1024 & 1024 & 2048 & 1024 & 2048 \\
    RoPE theta
      & 10,000 & 10,000 & 10,000 & 10,000 & 50,000 \\
    RMS norm epsilon
      & $1 \times 10^{-6}$ & $1 \times 10^{-6}$
      & $1 \times 10^{-6}$ & $1 \times 10^{-6}$
      & $1 \times 10^{-6}$ \\
    Dropout
      & 0.2 & 0.2 & 0.2 & 0.2 & 0.2 \\
    \midrule
    \multicolumn{6}{l}{\textbf{Training hyperparameters}} \\
    Epochs
      & 20 & 20 & 40 & 20 & 60 \\
    Accumulated batch size per device
      & 128 & 128 & 128 & 128 & 128 \\
    Learning rate
      & $4 \times 10^{-5}$ & $4 \times 10^{-5}$
      & $2 \times 10^{-5}$ & $4 \times 10^{-5}$
      & $2 \times 10^{-5}$ \\
    Weight decay
      & 0.1 & 0.1 & 0.1 & 0.1 & 0.1 \\
    Warmup ratio
      & 0.1 & 0.1 & 0.1 & 0.1 & 0.1 \\
    \bottomrule
  \end{tabular}
  \end{small}
\end{table}


\begin{table}[t]

  \centering

  \small

  \begin{tabular}{lccc}

    \toprule

     & \textbf{12 qubits} & \textbf{14 qubits} & \textbf{16 qubits} \\

    \midrule

    Number of stages & 3 & 5 & 5 \\

    \midrule

    \multicolumn{4}{l}{\textbf{RL hyperparameters}} \\

    \midrule

    Batch size & 16 & 16 & 16 \\

    Epochs & $\left[2,3,4\right]$ & 2 & 2 \\

    Generations & 16 & 16 & 16 \\

    Learning rate & $\left[ 4,3,2 \right]\times 10^{-6}$ &  $\left[ 4,2,2,1,1 \right]\times 10^{-6}$ & $\left[ 4,2,1,1,1 \right]\times 10^{-6}$ \\

    Warmup ratio & 0.1 & 0.1 & 0.1 \\

    $\beta$ & 0.28 & 0.28 & 0.28 \\

    $\epsilon_\text{high}$ & 0.001 & 0.001 & 0.001 \\

    Top-$p$ & 0.95 & 0.95 & 0.95 \\

    \midrule

    \multicolumn{4}{l}{\textbf{Distillation hyperparameters}} \\

    \midrule

    Learning rate & $4\times10^{-5}$ & $4\times10^{-5}$ & $4\times10^{-5}$ \\

    Batch size & 512 & 512 & 512 \\

    Epochs & 5 & 5 & 10 \\

    Warmup ratio & 0.05 & 0.05 & 0.05 \\

    Weight decay & 0.1 & 0.1 & 0.1 \\

    Threshold & $\left[ 5,3.4,1.6 \right]\times 10^{-3}$ & $\left[ 5,3.4,1.6, 0.8, 0.4 \right]\times 10^{-3}$ & $\left[ 2,1.75,1.68,1.6,1.52 \right]\times 10^{-2}$ \\

    Top-$n$ & $\left[ 8,4,2 \right]$ & $\left[ 8,8,4,4,2 \right]$ & $\left[ 8,4,4,4,4 \right]$ \\

    \bottomrule

  \end{tabular}

  \caption{\textbf{Post-training hyperparameters for each qubit count.} Numbers separated by commas indicate that they were different for successive iterations of fine-tuning.}
  \label{tab:rl_hparams}
\end{table}

\subsection{Tabulated Results}
\label{appendix:tab_results}

In this section we tabulate numerical values for results shown in the main text. 

\subsubsection{Circuit generation accuracy}

Tables \ref{tab:main_results-Nemotron},~\ref{tab:main_results_gemma-pretrain} and \ref{tab:main_results_gemma-finetune} show tabulated values for energy accuracy results provided in the main text for Nemotron and Gemma models. Table~\ref{tab:pearson_correlations} tabulates the Pearson correlation coefficients between model accuracy and circuit length. 

\begin{table*}[ht]
\centering
\caption{\textbf{Circuit generation performance of the Nemotron model after supervised fine-tuning.}}
\label{tab:main_results-Nemotron}
\resizebox{\textwidth}{!}{%
\begin{tabular}{lccccccc}
\toprule
\textbf{Dataset} & \multicolumn{2}{c}{\textbf{vs.\ ADAPT (mHa)}} & \textbf{Fraction} & \multicolumn{4}{c}{\textbf{vs.\ CASCI (mHa)}} \\
\cmidrule(lr){2-3} \cmidrule(lr){5-8}
 & \textbf{Mean} & \textbf{Std} & \textbf{(\%)} & \textbf{Mean} & \textbf{Std} & \textbf{MAE} & \textbf{RMSE} \\
\midrule
\multicolumn{8}{l}{\textit{Best-of-16 generations}} \\
\midrule
12q, $\varepsilon=10$~mHa   & 0.19 & 1.83 & 73.8 & 9.72 & 1.98 & 9.72 & 9.91\\
12q, $\varepsilon=5$~mHa  & 1.37 & 1.89 & 49.5 & 6.21 & 1.91 & 6.21 & 6.50 \\
14q, $\varepsilon=16$~mHa & 1.85 & 3.51 & 48.5 & 17.44 & 3.65 & 17.45 & 17.83 \\
14q, $\varepsilon=5$~mHa  & 9.34 & 6.17 & 7.8 & 6.7 & 6.17 & 14.26 & 15.54\\
16q, $\varepsilon=15$~mHa & 15.78 & 11.36 & 9.3 & 30.62 & 11.39 & 30.62 & 32.67\\
\midrule
\multicolumn{8}{l}{\textit{Increased training data}} \\
\midrule
12q, $\varepsilon=5$~mHa (+37\%) & 0.67 & 1.45 & 65.5 & 5.51 & 1.48 & 5.51 & 5.71 \\
14q, $\varepsilon=16$~mHa (+21\%) & 1.20 & 2.89 & 55.4 & 16.82 & 3.02 & 16.82 & 17.08 \\
\bottomrule
\end{tabular}
}
\end{table*}

\begin{table*}[ht]
\centering
\caption{\textbf{Circuit generation performance of the Gemma model after pretraining only.}}
\label{tab:main_results_gemma-pretrain}
\resizebox{\textwidth}{!}{%
\begin{tabular}{lccccccc}
\toprule
\textbf{Dataset} & \multicolumn{2}{c}{\textbf{vs.\ ADAPT (mHa)}} & \textbf{Fraction} & \multicolumn{4}{c}{\textbf{vs.\ CASCI (mHa)}} \\
\cmidrule(lr){2-3} \cmidrule(lr){5-8}
 & \textbf{Mean} & \textbf{Std} & \textbf{(\%)} & \textbf{Mean} & \textbf{Std} & \textbf{MAE} & \textbf{RMSE} \\
\midrule
\multicolumn{8}{l}{\textit{Best-of-16 generations}} \\
\midrule
12q, $\varepsilon=10$~mHa   & -0.13 & 1.09 & 87.4 & 9.40 & 1.24 & 9.40 & 9.48 \\
12q, $\varepsilon=5$~mHa  & 0.77 & 3.71 & 68.6 & 5.62 & 3.72 & 5.62 & 6.73 \\
14q, $\varepsilon=16$~mHa & 0.62 & 1.62 & 67.6 & 16.22 & 1.74 & 16.22 & 16.31  \\
14q, $\varepsilon=5$~mHa  & 4.59 & 2.58 & 7.8 & 9.51 & 2.58 & 9.51  & 9.85 \\
16q, $\varepsilon=15$~mHa & 7.56 & 5.31 & 11.9 & 22.41 & 5.34 & 22.41 & 23.03 \\
\midrule
\multicolumn{8}{l}{\textit{Increased training data}} \\
\midrule
12q, $\varepsilon=5$~mHa (+37\%) & 0.49 & 0.94 & 76.4 & 5.33 & 0.96  & 5.33 & 5.42 \\
14q, $\varepsilon=16$~mHa (+21\%) & 0.41 & 1.93 & 69.3 & 16.03 & 1.89 & 16.0 & 16.1 \\
\bottomrule
\end{tabular}
}
\end{table*}

\begin{table*}[ht]
\centering
\caption{\textbf{Circuit generation performance of the Gemma model after RL fine-tuning and distillation.}}
\label{tab:main_results_gemma-finetune}
\resizebox{\textwidth}{!}{%
\begin{tabular}{lccccccc}
\toprule
\textbf{Dataset} & \multicolumn{2}{c}{\textbf{vs.\ ADAPT (mHa)}} & \textbf{Fraction} & \multicolumn{4}{c}{\textbf{vs.\ CASCI (mHa)}} \\
\cmidrule(lr){2-3} \cmidrule(lr){5-8}
 & \textbf{Mean} & \textbf{Std} & \textbf{(\%)} & \textbf{Mean} & \textbf{Std} & \textbf{MAE} & \textbf{RMSE} \\
\midrule
\multicolumn{8}{l}{\textit{Best-of-16 generations}} \\
\midrule
12q, $\varepsilon=10$~mHa   & -5.92 & 1.32 & 99.68 & 3.61 & 1.30 & 3.61 & 3.83 \\
12q, $\varepsilon=5$~mHa  & -2.31 & 1.08 & 99.3 & 2.53 & 1.08 & 2.53 & 2.75 \\
14q, $\varepsilon=16$~mHa & -6.69 & 2.84 & 99.1 & 8.91 & 2.88 & 8.91 & 9.36 \\
14q, $\varepsilon=5$~mHa  & 2.86 & 2.17 & 17.2 & 7.78 & 2.17 & 7.78 & 8.08  \\
16q, $\varepsilon=15$~mHa & 5.57 & 5.39 & 22.2 & 20.4 & 5.42 & 20.4 & 21.6 \\
\midrule
\multicolumn{8}{l}{\textit{Increased training data}} \\
\midrule
12q, $\varepsilon=5$~mHa (+37\%) & -2.81 & 0.73 & 99.5 & 2.04 &  0.73 & 2.04 & 2.16 \\
14q, $\varepsilon=16$~mHa (+21\%) & -7.49 & 2.61 & 99.5 & 8.13 &  2.61 & 8.13 & 8.53 \\
\bottomrule
\end{tabular}
}
\end{table*}

\begin{table}[t]
    \centering
    \caption{
        \textbf{Pearson correlation coefficients between the reference energy and 
        the number of generated fermionic operators, for each dataset and model.}
    }
    \label{tab:pearson_correlations}
    \begin{tabular}{lccc}
        \hline
        Dataset & Nemotron SFT & Gemma Pretraining & Gemma RL Fine-tuning \\
        \hline
        12q, $\varepsilon=10$~mHa  & 0.73 & 0.39 & 0.29 \\
        12q, $\varepsilon=5$~mHa & 0.67 & 0.03 & 0.32 \\
        14q, $\varepsilon=16$~mHa  & 0.59 & 0.36 & -0.17 \\
        14q, $\varepsilon=5$~mHa & 0.39 & 0.55 & -0.11 \\
        16q, $\varepsilon=15$~mHa        & 0.58 & 0.72 & -0.20 \\
        \hline
    \end{tabular}
\end{table}

\subsubsection{Compute time comparisons}
\label{appendix:compute}

In Table~\ref{tab:ADAPT-VQE-scaling}, we tabulate the runtime for a single ADAPT-VQE iteration across our different problem sizes (12, 14, 16 qubits) using the UCCGSD operator pool. 
In Table~\ref{tab:model_speedup} we tabulate results for the inference and energy evaluation time for both models, compared to circuit generation with ADAPT-VQE. For ADAPT-GQE results we generate 16 candidate circuits per molecule 
and evaluate all 16 energies on a single NVIDIA H100 GPU using CUDA-Q (FP64 statevector simulation). Circuit generation is performed on a single GB200 node (4$\times$GB200, 186\,GB each, NVLink~5, 1.8\,TB/s) for the fine-tuned Nemotron-H 12B model and on an H100 for the Gemma model. The speedup is defined as the total time for ADAPT-VQE relative to the total time for both circuit generation and energy evaluation, $T_\text{ADAPT-VQE} / (T_\text{gen} + T_\text{eval})$.

\begin{table}[t]
  \centering
  \captionsetup{width=\linewidth, justification=raggedright, singlelinecheck=false}
  \caption{\textbf{Runtime of a single ADAPT-VQE iteration using the UCCGSD operator pool for 12-, 14-, and 16-qubit systems.}
  Results are reported for execution on a single CPU core, a single NVIDIA H100 GPU, and eight H100 GPUs using CUDA-Q.}
  \label{tab:ADAPT-VQE-scaling}

  \begin{tabular*}{\linewidth}{@{\extracolsep{\fill}}l
      S[table-format=5.2]
      S[table-format=3.2]
      S[table-format=3.2]
    @{}}
    \toprule
    \textbf{Qubits} & {\textbf{1 CPU (s)}} & {\textbf{1 GPU (s)}} & {\textbf{8 GPUs (s)}} \\
    \midrule
    12 & 10153.72 &  43.39 &   6.39 \\
    14 & \multicolumn{1}{c}{--} & 277.53 &  91.09 \\
    16 & \multicolumn{1}{c}{--} & 718.70 & 152.99 \\
    \bottomrule
  \end{tabular*}
\end{table}

\begin{table*}[ht]
\centering
\small
\caption{
\textbf{Computational cost comparison between model-generated circuits and circuit generation with ADAPT-VQE.} 
}
\label{tab:model_speedup}

\resizebox{\textwidth}{!}{%
\begin{tabular}{llccc}
\toprule
\textbf{Model} & \textbf{Metric}
& \textbf{12q ($\varepsilon{=}5$ mHa)}
& \textbf{14q ($\varepsilon{=}5$ mHa)}
& \textbf{16q ($\varepsilon{=}15$ mHa)} \\
\midrule

\multirow{8}{*}{Nemotron-H 12B}
& $T_{\rm gen}$ total (s)        & 164.00 & 293.30 & 563.90 \\
& $T_{\rm gen}$/circuit (s)      & 10.25  & 18.33  & 35.25  \\
& $T_{\rm eval}$ total (s)       & 1.41   & 1.67   & 1.20   \\
& $T_{\rm eval}$/circuit (s)     & 0.088  & 0.105  & 0.076  \\
& $T_{\rm ADAPT\mbox{-}VQE}$ (s) & 2,333.0 & 25,341.3   & 23,404.5 \\
& Speedup                       & $14.1\times$ & $85.9\times$ & $41.4\times$ \\
\midrule

\multirow{8}{*}{Gemma}
& $T_{\rm gen}$ total (s)        & 1.86 & 3.66 & 1.62 \\
& $T_{\rm gen}$/circuit (s)      & 0.12 & 0.22 & 0.10 \\
& $T_{\rm eval}$ total (s)       & 0.65 & 0.62 & 1.02 \\
& $T_{\rm eval}$/circuit (s)     & 0.041 & 0.039 & 0.064 \\
& $T_{\rm ADAPT\mbox{-}VQE}$ (s) & 2,333.0 & 25,341.3 & 23,404.5 \\
& Speedup                       &  $\bm{929.5\times}$ & $\bm{5,920.9\times}$ & $\bm{8,865.3\times}$  \\

\bottomrule
\end{tabular}%
}
\end{table*}

\end{document}